\documentclass[useAMS,usenatbib,usegraphicx,epsfig,subfigure]{mn2e}
\usepackage{amssymb,amsmath,supertabular}
\title[]{ Detection of dark galaxies and circum-galactic filaments fluorescently illuminated by a quasar at
  z=2.4\thanks{Based on observations collected at the European
  Organisation for Astronomical Research in the Southern Hemisphere,
  Chile under program 085.A-0171A.} }
\author[]{Sebastiano Cantalupo$^{1,2}$\thanks{E-mail:cantal@ucolick.org}, Simon J. Lilly$^3$ and Martin G. Haehnelt$^{1,4}$\\
$^1$ Kavli Institute for Cosmology, Cambridge \& Institute of
Astronomy, Madingley Road, Cambridge CB3 0HA, UK.\\
$^2$ Department of Astronomy and Astrophysics \& UCO/Lick Observatory, University of
California, Santa Cruz, CA 95064, USA.\\
$^3$ Institute for Astronomy, ETH Zurich, Wolfgang-Pauli-Strasse 27,
CH-8093 Zurich, Switzerland.\\
$^4$ Kavli Institute for Theoretical Physics, University of
California, Santa Barbara, CA 93106, USA.}

\begin{document}

\date{Accepted 2012 June 15. Received 2012 June 10; in original form
  2012 April 24}
\pagerange{\pageref{firstpage}--\pageref{lastpage}} \pubyear{}

\maketitle

\label{firstpage}

\begin{abstract}

A deep narrow-band survey for Ly$\alpha$ emission carried out
on the VLT-FORS
has revealed 98 Ly$\alpha$ candidates down
to a flux limit of $F_{\mathrm{Ly\alpha}}\sim 4\times 10^{-18}$ erg
s$^{-1}$ cm$^{-2}$
in a volume of 5500 comoving Mpc$^{3}$ at $z=2.4$
centered on the hyperluminous quasar HE0109-3518. 
The properties of the detected sources 
in terms of their
i) equivalent width distribution,
ii) luminosity function, and iii) the average 
luminosity versus projected distance from the quasar,
all suggest that a large fraction of these objects have been
fluorescently ``illuminated'' by HE0109-3518. 
This conclusion is
supported by comparison with 
detailed radiative transfer
simulations of the effects of the quasar illumination. We therefore have a unique opportunity to directly detect and image 
in emission dense gas, independent of any associated
star formation activity.  18 objects,
a much larger fraction than in ``blank-field'' Ly$\alpha$ surveys at
similar redshifts, have a rest-frame Equivalent Width (EW$_0$) larger than $240\mathrm{\AA}$, the expected 
limit for Ly$\alpha$ emission powered by Population II star
formation. 12 sources among these do not have any continuum counterpart in a
deep V-band imaging of the same field (reaching to $V(1\sigma)\sim30.3$ AB). For these, a
stacking analysis indicates
EW$_0 > 800\mathrm{\AA}$ ($1\sigma$), effectively ruling out 
Ly$\alpha$ powered by internal star formation.      
These sources are thus the best candidates so far for proto-galactic
clouds or ``dark'' galaxies at high-redshift, whose existence has recently been suggested by
several theoretical studies. Assuming they are mostly ionized by the quasar
radiation, we estimate that their gas masses would be about $M_{\mathrm{gas}}\sim10^{9}
M_{\odot}$ implying that their star formation efficiencies (SFEs)
are less than $10^{-11}$ yr$^{-1}$, equivalent to a gas
consumption time of 100 Gyr, several times below the SFE of
the most gas-rich dwarf galaxies locally, and two hundred times lower than typical massive star-forming galaxies at $z \sim 2$.
We have also discovered extended, filamentary gas, also likely illuminated by
the quasar, around some of the brightest continuum-detected sources 
with EW$_0 > 240\mathrm{\AA}$.
The morphology and luminosity of this extended emission are compatible with
the expectations for circumgalactic cold streams, as 
predicted by recent numerical models, but other origins, including tidal stripping, are also possible.
\end{abstract}


\newcommand{\QSOtwo}{HE0109-3518}
\newcommand{\angs}{\mathrm{\AA }}

\section{Introduction}

The Intergalactic Medium (IGM) plays a crucial role in the formation and evolution
of the structures in the Universe. 
Galaxies form from the IGM that settles into 
gravitational potential wells deep enough for the gas to
collapse (Rees \& Ostriker 1977, White \& Rees
1978). Radiative cooling dissipates thermal energy 
allowing the gas to concentrate at the center of the
halo and to generally form a (dense) disk. Following fragmentation, star formation takes place. 
Most of the observational studies conducted so far have captured the galaxy formation 
process from this stage on. Despite their importance, the early phases 
which may involve potential wells too shallow to efficiently form stars
are largely unobserved at present and remain poorly constrained.
Key questions about how the IGM is converted into stars at high-redshift and how efficient
this process is remain unanswered.

Recent theoretical studies have suggested that gas-rich, low-mass 
haloes ($10^9-10^{11}$ M$_{\odot}$) at high redshift may have very low star formation 
efficiencies, as a consequence of lower metallicities (e.g., Gnedin \&
Kravtsov 2010, Krumholz \& Dekel 2011), H$_{2}$ self-regulation
effects (e.g., Kuhlen et al. 2012),
or reduced cooling accretion rates due to local sources (e.g., Cantalupo
2010). 
If their star-formation efficiencies are suppressed, these ``dark'' galaxies can maintain a reservoir of gas that will be available for later 
star-formation in more massive systems.  This phase of suspended activity may be
required in order to simultaneously account for several properties of the
current galaxy population, like the observed slopes of the SFR-mass and
the Tully-Fisher relation, as has been suggested by Bouch\'e et al. (2010).

The potential importance of this early phase of galaxy formation
raises the obvious question of 
how to discover and study such gas clouds that do not have significant star-formation.
One possibility is try to detect them in HI absorption against background
quasars (see, e.g., Rauch 1998 and Wolfe, Gawiser \& Prochaska 2005
for reviews). For instance, recent studies by Fumagalli, O'Meara \& Prochaska (2011) and Cooke et al. (2011) 
have revealed high column-density systems
that have properties which may be compatible with these proto-galactic clouds,
given their metallicity consistent with primordial or Population III stars.
Unfortunately, these one-dimensional studies cannot
directly discriminate between truly isolated clouds
or components of larger galactic reservoirs, either around or within galaxies. 
Detecting these systems in emission rather than in absorption
would provide crucial information about their sizes, morphologies
and masses, potentially leading to a direct constraint on their
star formation efficiencies.

Direct imaging of HI in the 21cm line is unfortunately restricted to 
the local Universe (e.g. Giovanelli et al. 2005), given the intrinsic faintness of this atomic 
transition. Observational attempts to detect ``dark'' galaxies with this
technique in the local Universe have not so far produced positive results (e.g. Gavazzi et al. 2008). A significant
fraction of local dwarf galaxies do show, however, a much lower star formation
efficiency than is found in more massive spiral galaxies (Geha et al. 2006,
Schiminovich et al. 2010).  Extending
these direct imaging studies to high-redshift, where the cosmic 
star formation rate peaks, may require however an intrinsically brighter tracer of cosmic
hydrogen: i.e. the HI Ly$\alpha$ line. 

It has been predicted for many years (e.g., Hogan \& Weymann 1987) 
that the neutral gas that is responsible for quasar 
absorption lines at high redshift should produce potentially detectable fluorescent Ly$\alpha$ emission at 1216$\mathrm{\AA}$
from the incident 
ionizing radiation field at $\lambda<912\mathrm{\AA}$.  
For regions that are optically thick to ionizing radiation 
and are isotropically illuminated from outside,
a thin layer around the outer surface should act as a simple fluorescent 
mirror, converting up to 60\% of the incident ionizing radiation 
into Ly$\alpha$ photons (Gould \& Weinberg 1996).  Velocity field and geometrical effects can reduce 
the signal by half (Cantalupo et al. 2005).  
Unfortunately, the cosmic UV background is rather weak - 
whether predicted theoretically (e.g., Haardt \& Madau 2012)
or estimated observationally using the 
proximity effect in quasar absorption spectra (e.g., Bajtlik, Duncan \& Ostriker
1988, Rauch et
al. 1997, Scott et al., 2002, Bolton et al. 2005, Dall'Aglio et
al. 2008, D'Odorico et al. 2008).
This makes the detection of 
fluorescent emission from optically thick clouds in the field, i.e. illuminated by the general cosmic UV background,
extremely challenging, and perhaps impossible, even on 8-10 m 
class telescopes (see, e.g., Rauch et al. 2008).  
However, local enhancements in the ionizing background can be used to substantially 
increase the signal (e.g., Haiman \& Rees 2001). For instance, gas in the vicinity of a bright
quasar may be exposed to a stronger UV flux compared with to an ``average'' cloud, and are then expected 
to be correspondingly brighter in fluorescent Ly$\alpha$ (e.g., Cantalupo et al 2005).

The detection of the line-of-sight proximity effect in quasar spectra provides
evidence - in absorption - that bright quasars are indeed
``illuminating'' and ionizing gas clouds within several comoving Mpc scales, at least along our line-of-sight. 
Searches for a similar effect along the transverse
direction using projected pair of quasars, the so called ``transverse proximity-effect'', 
have however produced mixed results so far (e.g., Jakobsen et al. 2003, 
Hennawi \& Prochaska 2007, Worseck et al. 2007, Goncalves, Steidel \&
Pettini 2008, Kirkman \& Tytler 2008), indicating possible
emission anisotropies or short ages for the quasar
bright phase ($<1$ Myr). Detection of quasar
fluorescence - the ``counterpart'' in emission of the proximity effect -  
may thus provide direct constraints on angular emission
properties and lifetimes of these sources.

\subsection{Previous observational attempts to detect fluorescent emission near quasars}

 In the last years, we have seen an increase of our knowledge of 
the quasar fluorescence process with detailed numerical studies (e.g.,
Cantalupo et al. 2005, Kollmeier et al. 2010).
However, until very recently, dedicated observational surveys have been
limited to a few, generally unsuccessful, searches. 
The first fluorescent survey around a quasar, based on the simple expectations from 
the slab model of Gould \& Weinberg (1996) was
performed by Francis \& Bland-Hawthorn (2004)
using a tunable filter imager on the 4m Anglo-Australian Telescope.  This produced a null result. 
However, as demonstrated by Cantalupo et al. 2005 (see also
the discussion in Cantalupo et al. 2007), this null result is in fact quite consistent
with theoretical expectations from quasar illumination 
considering: i) a more realistic density distribution than simple slab geometry 
for the gas, ii) the proximity effect of the quasar itself, which reduces
the sizes and number density of optically thick clouds\begin{footnote}{These effects explained
the null detection of fluorescent sources. However, it is still unclear why no
line emitting galaxies were detected in the same field (when at least
6 Ly$\alpha$ emitters were expected). 
}\end{footnote}.
A few years later, Adelberger et al. (2006) discovered, serendipitously, a double peaked
Ly$\alpha$ emitter associated with a Damped Ly$\alpha$ absorbing system in the proximity of
a bright quasar. The Ly$\alpha$ profile was consistent with expectations
from fluorescence and indicated that the gas was certainly optically thick. However, this
objects was also associated with a star forming galaxy that was bright enough in the continuum to entirely
power - internally - the observed Ly$\alpha$ emission, given its rest-frame
Equivalent Width (EW$_0$) of
about 75$\mathrm{\AA}$, well within the range that can be produced by
conventional stellar populations
(e.g., Schaerer 2002). 
A systematic survey for other similar systems in quasar pairs
by Hennawi et al. (2006) has until now yielded only one Ly$\alpha$ emitting source (Hennawi et al. 2009).  However, these authors concluded that this object
is likely intrinsic to the quasar host and similar to the Ly$\alpha$ fuzz detected around many active
galactic nuclei (e.g., Heckman et al. 1991; see also Francis \& McDonnell 2006).

Motivated by our own more sophisticated theoretical modeling, in
Cantalupo et al. (2007) we carried out a blind
spectroscopic survey 
around a bright quasar at $z = 3.1$ using the VLT/FORS2 instrument in
a ``multi-slit plus filter'' 
mode (Crampton \& Lilly 1999), using two OIII narrow band filters.  
This method allowed us to perform a deep spectroscopic survey down to
a flux limit of $F_{ly\alpha}\sim2\times10^{-18}$ erg s$^{-1}$ cm$^{-2}$
over a significant but non-contiguous volume around
the quasar ($\sim1700$ comoving Mpc$^3$).
In good agreement with theoretical expectations, we found 13 line
emitters consistent with Ly$\alpha$, a third of which 
showed signatures of the double-peaked profiles expected from fluorescence. 
Only 2 of these 13 objects had a significant ($2\sigma$)
detection of underlying stellar continuum emission 
in a 2 hour deep V-band image of the same field. 
Unfortunately, the depth of the V-band image was not sufficient to rule out
the presence of internal ``illumination'' by star formation as the source of the ionizing radiation.  However, other indications, including the 
constraints on the Ly$\alpha$ surface brightness as a function of
distance from the quasar, suggested that about
half of the sample could have been indeed fluorescently illuminated by the quasar. 

A blind spectroscopic survey has the advantage of
reducing the sky-background and giving immediate confirmation 
of the line-emitting nature of the source.
However, there are also several limitations of this approach, including: 
i) a reduction in the
sampled volume, by a factor $\sim14$ for our ``multi-slit plus
filter'' configuration, compared to narrow-band imaging on the same
instrument, ii) an inability to detect extended objects,
e.g. filamentary structure, with sizes larger than the slit width
(2'' in our case), iii) the absence of unambiguous
information on the sizes and positions of objects (since one spatial dimension is effectively lost), which are crucial to better
constrain the EW$_0$ by comparison with a broad-band image.

In this paper, we present a new survey that is based on deep
narrow-band (NB) imaging on VLT-FORS that overcomes these limitations and
provides the first large statistical sample of cosmic gas clouds
that have been fluorescently illuminated by a quasar. 

The layout of the paper is as follows. In Section \ref{datasec}, we present
the survey design, observations, data reduction and Ly$\alpha$ candidate selection.
In Section \ref{resultsec}, we show the observational results. In
Section \ref{originsec}, we demonstrate that many of the objects detected have likely been fluorescently illuminated by the quasar. In particular, we compare the EW
distribution and Ly$\alpha$ luminosity
function with those from field surveys, and examine the relation between object luminosity and the projected distance from
the quasar.  In Section \ref{naturesec}, we discuss the nature and physical properties of this new
population of Ly$\alpha$ sources and the implications of
our identification as fluorescent emission. In Section \ref{uncertaintiessec}, we  
discuss the uncertainties and limitations of our survey. We
summarize and conclude the paper in Section \ref{summarysec}.
Through the paper, we use a ``standard'' $\Lambda$CDM cosmology with
$H_0=71$ km s$^{-1}$ Mpc$^{-1}$, $\Omega_{m}=0.27$ and $\Omega_{\Lambda}=0.73$
from WMAP seven-year release (Komatsu et al. 2011).

\section{The data}\label{datasec}

\subsection{Survey Design}

Our survey design has been motivated by two simple but crucial
requirements: i) to maximize the fluorescently illuminated volume, by 
selecting one of the brightest quasars in the sky,  ii) to minimize the
sky-background by having the narrowest bandwidth filter that contains
such a volume.  In order to meet these two requirements,
it is vital that the systemic redshift of the quasar is known with good
accuracy, e.g. from low-ionization lines or narrow forbidden lines
such as [OIII]. Unfortunately, not many quasars have such a measure
- and our previously observed quasar at $z=3.1$ is not one of these -
especially if we also require them to be within an existing
NB filter. Although we had already several detections around the
$z=3.1$ quasar, a large fraction of these would have been too faint to be
re-detected in NB imaging.  For the above reasons, we designed a new custom NB filter to take advantage of the new
blue sensitive E2V detector on FORS and
enabling us to select an
ultraluminous quasar at $z=2.4$.  Due to the lower surface brightness redshift dimming -
a factor of about 2 - and the increased throughput, an imaging survey based on a 20
hour NB image at $z=2.4$ should reach a similar depth to our previous
spectroscopic survey at $z=3.1$. 
As we will show, the much larger volume probed by NB
imaging results in a substantially larger sample of Ly$\alpha$
sources - about a hundred - allowing us to perform crucial statistical
analyses to confirm the fluorescent nature of these sources,
 e.g. the distribution of Ly$\alpha$ EW 
compared with similar surveys that did not target bright quasars.   
Moreover, we are now able to constrain the sizes and masses of these
systems and to detect extended, circumgalactic emission. 

\subsection{Observations and data reduction}

Observations were taken during four visitor-mode nights at the VLT 8.2m telescope Antu (UT1)
on 2010 September 9-12 using the FORS instrument in imaging mode with the blue
sensitive E2V CCD. A custom-built interference filter (FILT\_414\_4) with central wavelength 
$\lambda_{\mathrm{NB}}=414.5$nm and full-width-half-maximum FWHM$=4$nm has been used
to image the field of  
\QSOtwo\ ($z_{sys}=2.4057\pm0.0003$, $b_J=16.7$), sensitive to HI Ly$\alpha$ over a narrow ($\Delta z=0.033$) redshift
range around the quasar. The quasar systemic redshift 
is precisely constrained by the wavelength of the detected [OIII] $\lambda5007$ emission line 
(Shemmer et al. 2004, Marziani et al. 2009). The quasar is radio-quiet.
The field was also observed in the B$_{\mathrm{HIGH}}$ and V$_{\mathrm{HIGH}}$ filters. 
A total of 20 hours integration through the narrow-band filter, 6 hours through the V-band (of which 4.5 hours were used)
and 0.9 hours through the B-band filter were obtained. The observation log 
is reported in Table 1. 
Observations were split into separate exposures of 1200 s in the narrow-band and
120-300 s in the broad bands. Individual exposures were shifted and rotated with respect to each other
using a semi-random shifting pattern of radius 15'' and four different position angles (0, 90, 180 and 270 degrees) 
in order to facilitate the removal of cosmic rays, CCD cosmetic, ghosts and residual
flat-field errors. 
The first part of the night beginning 2010 September 12 suffered from poor seeing conditions,
affecting 18 V-band exposures (for a total of 5400s). These have been discarded before image combination. 
Therefore, the final V image corresponds to a combined integration time of 4.5 hours. 

 The images have been reduced using standard routines within the reduction software IRAF, including 
bias subtraction, flat fielding, CCD illumination correction and image combination. 
A combination of twilight sky flats
and unregistered science frames has been used to produce flat field images and illumination
corrections for each band and position angle. 
Sky subtraction was performed, image by image, fitting a polynomial surface function of order
two in order to compensate for small residuals and possible sky variation. Before image combination,
each image was registered and corrected for distortion and rotation with the IRAF tasks ``geomap''
and ``geotran'' using a series of unsaturated stars and a polynomial surface fit of order 3.
Finally, for each band, the corrected frames were combined with an averaged sigma-clipping
algorithm. 

The combined science images have been registered on the ICRF frame of reference using 
the USNO-B1 catalogue. The uncertainty of $0''.2$ in the USNO-B1 catalogue
dominates the astrometric error, except at the edge of the field of view,
where residuals from distortion correction increase the error to about $0''.5$. 
The photometric calibration was performed using several
spectrophotometric standard stars from the catalogue of Oke (1990). For the calibration
of the NB image we have chosen three standard stars (LDS749B, LTT9491, NGC7293) 
that did not present any spectral feature in the relevant wavelength range. The
derived zero points were consistent with each other within 
few percent.  Magnitudes quoted here are AB magnitudes
corrected for galactic extinction 
(0.09 magnitudes for NB and B band
and 0.07 magnitudes for the V band; Schlegel et al. 1998).
Given the large number of exposures
of the same field at different air-masses we were able to determine
the atmospheric extinction coefficients for each night using relatively bright,
unsaturated stars.

\begin{center}
\begin{table}
\begin{tabular}{l l c c}
\multicolumn{4}{c}{\textbf{Table 1}: \QSOtwo\ field observation log} \\
\hline
\hline
\multicolumn{1}{c}{Date} &
\multicolumn{1}{c}{Filter} &
\multicolumn{1}{c}{Seeing$^a$} &
\multicolumn{1}{c}{t$_{\mathrm{exp}}$(s)$^b$} \\
\hline
2010 Sep. 9 & V$_{\mathrm{HIGH}}$ & 0''.6 & 5400 \\
            & FILT\_414\_4        & 0''.7 & 19500 \\
            & B$_{\mathrm{HIGH}}$ & 0''.6 & 720 \\
2010 Sep.10 & V$_{\mathrm{HIGH}}$ & 0''.7 & 5400 \\
            & FILT\_414\_4        & 0''.7 & 21600 \\
            & B$_{\mathrm{HIGH}}$ & 0''.7 & 840 \\
2010 Sep.11 & V$_{\mathrm{HIGH}}$ & 0''.7 & 3600 \\
            & FILT\_414\_4        & 0''.7 & 11780 \\
            & B$_{\mathrm{HIGH}}$ & 0''.6 & 480 \\
2010 Sep.12 & V$_{\mathrm{HIGH}}$ & 1''.3$^c$/0''.8 & 7200 \\
            & FILT\_414\_4        & 0''.8 & 19200 \\
            & B$_{\mathrm{HIGH}}$ & 0''.7 & 1200 \\
\hline
TOTAL       & FILT\_414\_4        & 0''.7 & 72080 \\
            & V$_{\mathrm{HIGH}}$ & 0''.7$^d$ & 21600/16200$^d$ \\
            & B$_{\mathrm{HIGH}}$ & 0''.7 & 3240 \\
\hline
\multicolumn{4}{l}{$^a$ averaged value.} \\
\multicolumn{4}{l}{$^b$ total exposure time per night and filter.} \\
\multicolumn{4}{l}{$^c$ average value for the first exposures up to 5400s.} \\
\multicolumn{4}{l}{$^d$ discarding bad seeing ($>1''$) exposures.} \\
\end{tabular} 
\end{table}
\end{center}

\subsection{Ly$\alpha$ candidate selection}\label{LyaSel}

The selection and photometry 
of the Ly$\alpha$ candidates have been performed 
using an iterative method based on the
program SExtractor (Bertin \& Arnouts 1996; v2.5)
and IDL procedures.
The narrow-band (NB) image alone has been used for candidate detection.
Aperture photometry was subsequently performed on both
NB and broad-band images using SExtractor
dual-mode. This strategy 
was preferred above a combination of narrow and broad-band as
detection image (as sometimes done in the literature for
Ly$\alpha$ emitter searches, e.g., Fynbo et al. 2003)
since this would bias our results towards objects with
continuum-detection, while we are especially interested 
in detecting object without significant associated continuum.

In order to minimize spurious detections we varied the main
detection parameters in SExtractor creating a large number
of candidate catalogues. In particular, we used a minimum
area for detection (DETECT\_MINAREA) ranging from 4 to
8 pixels and a detection threshold (DETECT\_THRESH) 
spanning from 1.3 to 2 times the local noise level.
These parameter sets were applied to both unsmoothed
images and to images smoothed with two different gaussian
kernels.  Each SExtractor pass produced, on average, about 3000
detections.
The same procedure was applied to detecting candidates in
the broad-band images. These were associated by SExtractor,
whenever possible, to the objects found in the narrow-band 
using the dual-image mode.
Initial photometry was performed with isophotal apertures 
(FLUX\_ISO) and SExtractor FLUX\_AUTO for each band. 

As a first step in the selection of Ly$\alpha$ candidates,
we derived an initial guess on the candidate observed EWs
from the magnitudes obtained by SExtractor:
\begin{equation}
EW=\Delta\lambda_{\mathrm{NB}}\left[\frac{\lambda_{\mathrm{V}}}
{\lambda_{\mathrm{NB}}}\right]^{\beta_{\lambda}+2}\times 
(10^{-0.4(\mathrm{NB}_{\mathrm{AB}}-\mathrm{V}_{\mathrm{AB}})}-1). 
\end{equation}
Where $\Delta\lambda_{\mathrm{NB}}$ and $\lambda_{\mathrm{NB}}$ 
are, respectively, the NB filter 
FWHM and central wavelength,
$\lambda_{\mathrm{V}}$ is the V$_{\mathrm{HIGH}}$ filter effective
central wavelength, $\beta_{\lambda}$ is the slope of the UV continuum
(in units of erg s$^{-1}$ cm$^{-2}$ $\mathrm{\AA}$) 
between $\lambda_{\mathrm{V}}$ and $\lambda_{\mathrm{NB}}$,
$\mathrm{NB}_{\mathrm{AB}}$ and $\mathrm{V}_{\mathrm{AB}}$
are, respectively, the NB and V magnitude obtained by Sextractor.
We used the V-band to estimate the continuum since it is
not affected by the presence of the Ly$\alpha$ emission line.
With this strategy we are able to better select
objects without significant continuum emission. At the same time, this simplifies the
calculation of the EW.
When both are present, we used the V-band and B-band isophotal magnitude
to derive $\beta_{\lambda}$, otherwise we assigned 
to the source a flat continuum slope (in frequency), i.e. $\beta_{\lambda}=-2$.
This is the typical maximal value for a starburst galaxy
(e.g., Meurer et al. 1999). Ly$\alpha$
surveys at $2.1<z<2.3$ find $\beta_{\lambda}\sim-1.9$ for the most
luminous sources and higher values 
($\beta_{\lambda}\sim-1.5$ but with a large scatter)
for the faintest (e.g., Guaita et al. 2011,
Nilsson et al. 2009). Detected objects with $\beta_{\lambda}<-3$
are extremely rare. Note that a value of
$\beta_{\lambda}=-3$ would only imply a decrease
of about 25\% of our estimated EW.
For the objects without broad-band detection we
assigned an initial continuum 
flux corresponding to the average 
$2\sigma$ level within a 1'' photometric aperture
across the image (larger than the typical seeing of
$0''.7$). To convert the NB magnitude
to a line flux, we assumed that the emission line
is fully contained by the NB filter and centered
at the filter transmission peak\begin{footnote}{This is a similar
    approach with respect to the one used by other deep surveys not targeting a bright quasar 
    (e.g., Grove et al. 2009, Hayes et al. 2010) that we will use in
    section 4 for a direct comparison of the
    Ly$\alpha$ luminosity functions.}\end{footnote}. 

Since the filter
transmission curve has a roughly gaussian shape, the
derived flux is thus a lower limit to the actual emission. This implies that the quoted EW should also 
be considered as a lower limit, strengthening the
results discussed in section 3.
We did not attempt to remove the line contribution
from the B-band magnitude used to estimate $\beta_{\lambda}$.
This is also a conservative assumption for the
estimation of the EW.

For each catalogue we then performed a first
cut selecting objects with observed EW$>68\mathrm{\AA}$,
corresponding to a rest-frame EW$_0=\mathrm{EW}/(1+z) > 20\mathrm{\AA}$ for
Ly$\alpha$ at $z=2.4$. We chose this EW$_0$ cut
in order to be comparable with the majority
of Ly$\alpha$ surveys in the literature.
This produced Ly$\alpha$ candidate
catalogues of which the largest contained 224 objects. Then we cleaned
through visual inspection the spurious detections due to the
vicinity of bright stars, excessive fragmenting of objects
(e.g., galaxies showing multiple spatial components 
in the narrow-band and a single spatial component 
in the broad-band) and noise at the edges of the image.
This produced a set of catalogues containing a number of candidates 
varying from 89 to 100.

\begin{figure}
\begin{center}
\includegraphics[totalheight=0.35\textheight]{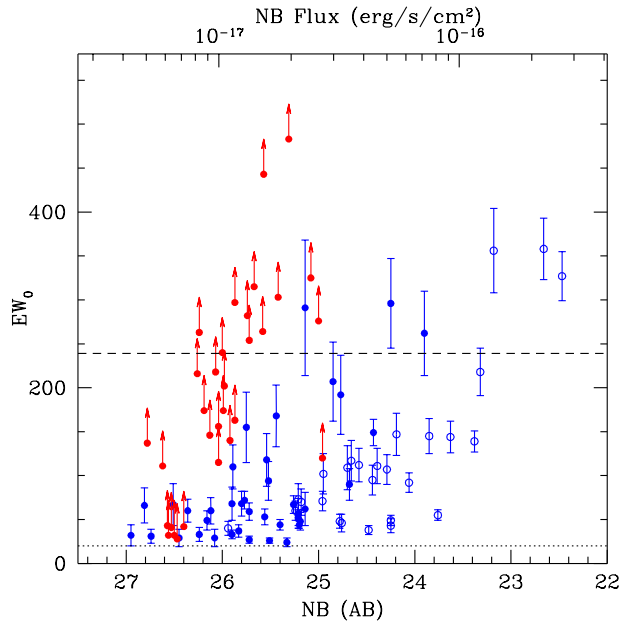}
\caption{Rest frame equivalent width (EW$_0$) of the detected objects 
versus their narrow-band magnitudes. Blue solid
(or open) circles represent candidates with V-band (or both V and B-band)
detections above the $3\sigma$ level for V-band (and $5\sigma$ for the B-band). When the B-band is detected (open circles) the EW$_0$ is derived from a continuum
conservatively estimated from both V and B-bands without correcting for the line contribution
to the B-band. For objects without B detection, we estimate the continuum from the V-band, assuming a continuum slope of $\beta_{\lambda}=-2$.
The red solid circles represent lower limits on the EW$_0$ of the
objects without continuum detection (i.e., V$ < 3\sigma$). These correspond to
the (local) 1$\sigma$ lower limits for objects with signal to noise ratio SNR$<0$
in V-band (the deepest available), otherwise the measured flux plus
the local 1$\sigma$ noise has been used (e.g., Feldman \& Cousins
1998; see text for details). 
The dashed line represents the theoretical limit on the EW$_0$ that is produced by ``normal''
stellar populations (EW$_0=240\angs \ $; e.g., Charlot \& Fall 1993, Schaerer 2002) while the dotted line is the lowest EW$_0$ for
candidate selection (EW$_0=20\angs \ $, consistent with most published Ly$\alpha$ surveys).
The continuum-undetected objects with EW$_0>240\mathrm{\AA}$ (red dots
above the dashed line) are the best candidates for proto-galactic, or
``dark'', gas-rich clouds that have been fluorescently illuminated by the quasar.
}  
\label{EW_NB}
\end{center}
\end{figure}

For the objects in the cleaned catalogues 
without Sextractor broad-band detections
we performed
again photometry using our own specifically designed 
IDL procedures in order to derive an upper limit on the
broad-band flux and therefore a lower limit on the EW.
The sources without Sextractor broad-band detection 
are typically spatially compact or unresolved 
in the NB image. Therefore we used circular apertures with
a fixed size of 1.5 arcsec (twice the seeing disk, 0.7'', that is 
coincidentally the same for each band) for both the
NB and broad-bands. Using exactly the same aperture for NB and
broad-band is again a conservative assumption 
for the calculation of the EW, 
since the gas extended distribution and scattering effects would 
likely result in a Ly$\alpha$ emission that is more extended than the stellar continuum (e.g., Cantalupo et
al. 2005; see also Steidel et al. 2011). The apertures were centered,
for all bands, at the location corresponding to the peak of the NB
emission (derived from SExtractor). We used relative astrometry to
refine the position of these apertures in the broad-bands. 
If the flux measured in the broad-band apertures was above 3
times the local noise level ($\sigma$) we considered the source
continuum-detected and we used the measured flux to compute the EW.
Otherwise, we consider the object undetected in the continuum and we
estimate an upper limit on the continuum flux using a procedure
similar to Feldman \& Cousins (1998). For apertures where 
the measured flux is positive this corresponds to
assigning an upper limit equal to the sum of the measured value
and the local 1$\sigma$. In case the flux is negative,
 although Feldman \& Cousins (1998) provide a table for
assigning upper limits (with values $<1\sigma$) in this situation,
we decided to use as a conservative value the local $1\sigma$ as a
minimum threshold. The continuum flux upper limits 
translates then into EW lower limits. With this method, we use the statistical
information contained in the measured flux and we have a consistent
measurement with the 1$\sigma$ errorbars of the detected objects,
i.e. we make sure that the actual EW is above the reported value within
\emph{at least} 1$\sigma$ confidence level. 
We note that current Ly$\alpha$ surveys in literature use different
methods to estimate continuum flux upper limits for undetected
sources. In particular, a common choice is to use a fix value (e.g., a
\emph{globally} estimated 1$\sigma$ or 2$\sigma$ value
for a given aperture) independently of the measured flux. This
complicates the comparison of different works in the literature.  

Bright foreground objects in proximity
to the aperture were masked, but we cannot rule out
the possibility that fainter foreground objects are contaminating
the apertures in the broad-band. Note that this would
imply that the derived EW of the objects in question are underestimated and thereby strengthen
the conclusions presented in the paper.
For the objects with SExtractor broad-band detection
we used FLUX\_AUTO to calculate total magnitudes and
isophotal apertures (FLUX\_ISO) to compute the colors and EW, as commonly
performed in previous Ly$\alpha$ searches (e.g., Hayes et al. 2010).
Isophotal apertures and FLUX\_AUTO measurement
did not produce magnitudes varying more than few percent
from each other, except in the few cases where 
sources were located close to foreground objects.  
Only objects with
a SNR$\gtrsim$5 in the narrow-band images were retained
in the catalogues. This corresponds to about
NB$_{\mathrm{AB}}(5\sigma)\sim26.8$ on average across
the image, while the typical (median) $1\sigma$ magnitude
limits for V-band and B-band were, respectively,
V$_{\mathrm{AB}}(1\sigma)\sim29.7$ and
B$_{\mathrm{AB}}(1\sigma)\sim29.1$ for a 1.5''
diameter circular aperture.  

In order to assess the level 
of spurious detections due to noise
for each set of parameters and filters we produced
a  \emph{negative catalogue} applying SExtractor to the
narrow-band image multiplied by $-1$. 
The number of detected objects in the negative catalogues 
varied from zero to 30. In the latter case, however,
these spurious detections were mostly confined at the
image edges, while only between three to five spurious objects were present
at the center. 
Among the eleven candidates that were not present in 
every \emph{positive catalogue}, nine did not have
any broad-band detection. For these nine candidates
we have rechecked the narrow-band image splitting the
total exposure time in four parts (each corresponding
to a given rotation angle). Six objects were detectable
(at SNR$>$3) in at least three out of four images. The 
remaining three objects, in accordance with the
number found in the negative catalogues, were considered
spurious and thus removed from their catalogues.
The final catalogue contained thus 97 sources.
In section \ref{fgSec}, we discuss in detail the possible
fraction of foreground contaminations in our sample.

Finally, we estimated the completeness of our NB imaging
with the standard procedure of distributing and recovering a large
number ($\sim1000$) of artificial sources with varying sizes.
In particular, we used the distribution of apertures derived by
SExactror for our detected candidates obtaining a completeness
factor $f_c=(25\%, 50\%, 75\%, 90\%)$ for $NB= 
(26.95, 26.55, 26.30, 26.12)$ corresponding to Ly$\alpha$ luminosities,
at $z=2.4$, $L_{\mathrm{Ly\alpha}}=(0.20, 0.30, 0.38, 0.43)$ in units of $10^{42}$ erg
s$^{-1}$. The factor $f_c$ will be used to correct the Ly$\alpha$
luminosity function for incompleteness in section \ref{LFsection}.

\begin{figure*}
\begin{center}
\includegraphics[totalheight=0.7\textheight]{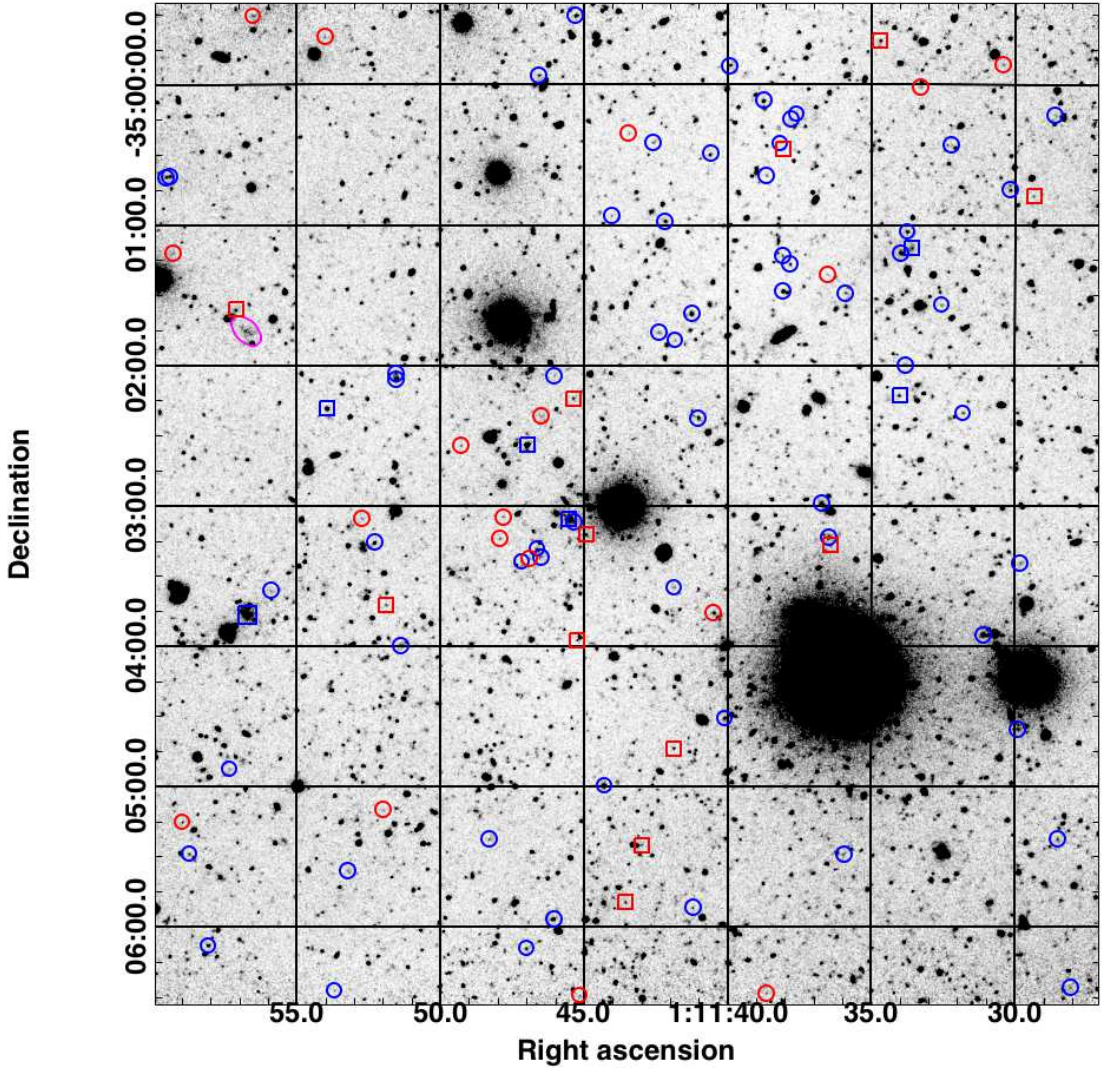}
\caption{The combined narrow-band image (20 hours total integration
  time) of the field surrounding
 \QSOtwo\ (near the center of the image). Squares (or circles) 
indicate the position of detected emission line objects with
EW$_0>240 \angs\ $ (or EW$_0<240\angs $), assuming the line is Ly$\alpha$.
Red (or blue) colors represent
objects undetected (or detected) in the broad-band images. The position of
the extended, 
blob-like emission source (LAB1) is indicated by
a magenta ellipse (approximatively at $ra=1:11:57$ and
$dec=-35:01:50$). 
Objects with undetected continuum and with
EW$_0 > 240\angs\ $ (i.e. the red squares) 
are the best candidates for
protogalactic clouds, or ``dark'' galaxies, fluorescently illuminated by the quasar
(see text for details).}
\label{NBimage}
\end{center}
\end{figure*}

\begin{figure*}
\begin{center}
\includegraphics[totalheight=0.8\textheight]{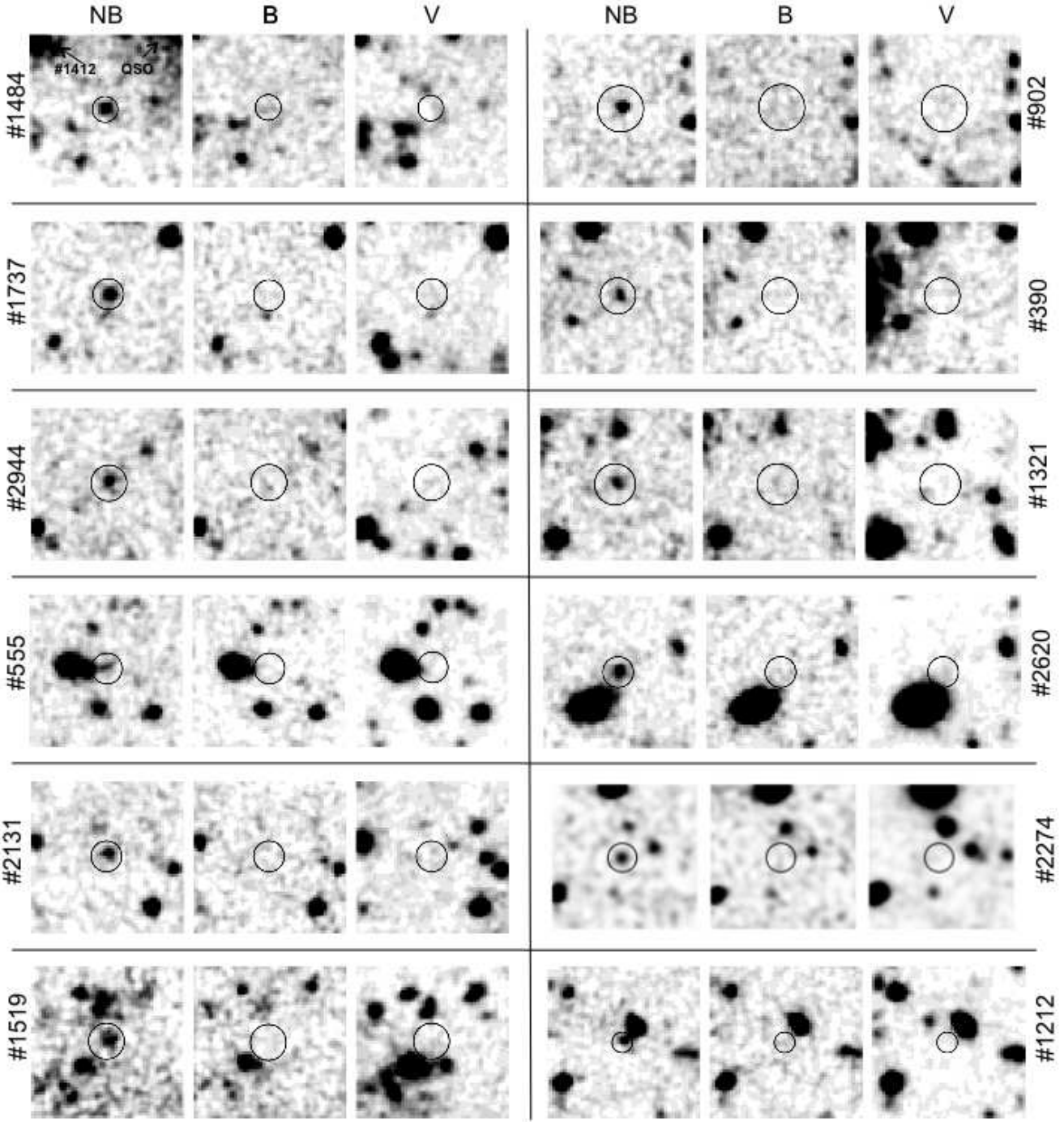}
\caption{Images of the narrow-band excess sources without continuum 
detection ($V<3\sigma$) and EW$_0>240 \angs\ $ (assuming the line is
Ly$\alpha$) in three different bands:
narrow-band (left-hand panels), B (central panels), and V (right-hand
panels). For an object at $z=2.4$, the central wavelength of these
filters correspond to the rest-frame
Ly$\alpha$, $1300\mathrm{\AA}$ and $1600\mathrm{\AA}$, respectively.
Note that the B-band contains also the source Ly$\alpha$ line, differently
from the V-band, although the emission would be typically too faint to be
detected in B.  
Panel dimensions are 15 $\times$ 15 arc sec$^2$. The images have been
smoothed with a gaussian kernel with radius $0".5$ (two pixels) for
clarity. These objects represent the
best candidates for fluorescent protogalactic clouds or ``dark'' galaxies
given the absence of detected continuum and the high lower limit on
the EW$_0$. As discussed in the text, an EW$_0 > 240 \angs\ $ strongly
suggests that internal star formation is not likely the origin of the 
Ly$\alpha$ emission and that we are detecting dense gas fluorescently
illuminated by the quasar.}
\label{PGCFig}
\end{center}
\end{figure*}

\begin{figure*}
\begin{center}
\includegraphics[totalheight=0.2\textheight]{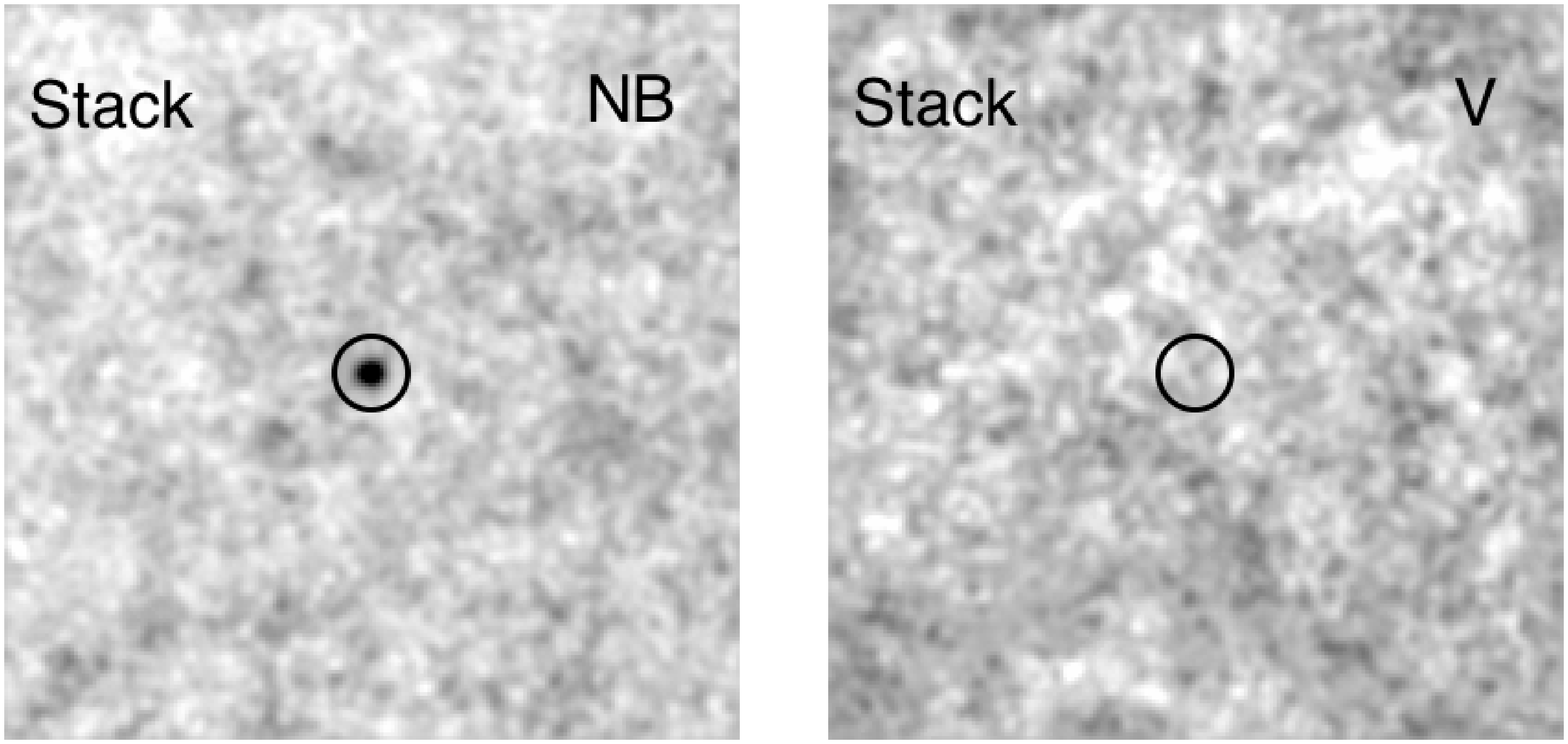}
\caption{Stacked NB image (left-hand panel) and V-band image
 (right-hand panel) of the 12 Ly$\alpha$ sources that did not have individual
 continuum detection and an implied EW$_0>240\mathrm{\AA}$.  The images have been
smoothed with a gaussian kernel with radius $0".5$ (two pixels) for
clarity. The V-band stack
does not show any continuum source above 1$\sigma$ level (V$\sim30.3$). The
combined constraint implies a EW$_0>800\mathrm{\AA}$, effectively
ruling out internal star-formation as the origin of the Ly$\alpha$ emission in these objects.}
\label{Stack}
\end{center}
\end{figure*}

\section{Results}\label{resultsec}

The final, merged catalogue consists of 97 narrow-band
excess objects. 
The position and full properties of all detected objects
are reported in tabular form in Tables \ref{PGCTable}, \ref{LAETable} and Appendix A. 
The EW (estimated as in section \ref{LyaSel} )
and NB magnitudes of these objects
are presented in Figure \ref{EW_NB}.
Among the 97 NB-detected objects, 31 do not have
any detectable continuum (above $3\sigma$) 
in the very deep broad-band
images (red symbols with arrows in Figure \ref{EW_NB})
and therefore the reported EW has to be considered
a lower limit as discussed in section \ref{LyaSel}. 
Given the sensitivity
of our broad-band imaging we are able to put
strong constraints on the EW limits, at least
for the objects with NB magnitudes NB$_{AB}<26$.
Among these, 12 Ly$\alpha$ candidates have a
lower limit EW$_0 > 240 \angs $, i.e. higher than the
typical maximum value produced by
normal stellar populations (e.g., Charlot \& Fall 1993,
Schaerer 2002; see also Malhotra \& Rhoads 2002 and
Raiter, Schaerer \& Fosbury 2010).
Six of the 66 objects that have at least one broad-band detection
(blue circles in Figure \ref{EW_NB})
also have  EW$_0 > 240 \angs $. As we will
show in this section, these contain also the objects with the
most evident signs of extended, filamentary emission.

In Figure \ref{NBimage}, we present the combined narrow-band image (20
hours combined integration time) with overlaid positions of
the 97 Ly$\alpha$ candidates. The quasar is near the
center of the image. Open squares (circles) represent
objects with a measured or lower limit 
EW$_0 > 240\angs\ $ (EW$_0 < 240\angs $)
while red (or blue) colors indicate if the object was
undetected (or detected) in any of the broad-band images.

From the image, we have visually identified 
an additional narrow-band excess source, i.e., 
a blob-like extended emission (magenta ellipse
in Figure \ref{NBimage} 
approximatively at $RA = 1:11:57$ and
$\delta = -35:01:50$). Due to the presence of relatively bright
components within this extended image, which appear also in the broad-band
(possibly, foreground objects), this extended source was split
into several parts by SExtractor and it did not make it
to the final catalogue using our selection technique. 
We will refer to this object as LAB1 from now on. 

In Figure \ref{PGCFig}, we show postage stamp images
of 12 candidates without continuum 
detection ($V<3\sigma$) and lower limit EW$_0>240\angs $.
The objects are ordered by decreasing EW limit
from the top to the bottom panels.
Their positions, photometric properties
and EW limits are reported in Table \ref{PGCTable}. Among the brightest candidates, 
\#1484 and \#902 have a very high lower limit
on their rest-frame Ly$\alpha$ equivalent widths,
namely EW$_0 >> 400 \angs $. Interestingly, \#1484
is also the source closest to the quasar (in projection), 
located just about 15 arcsec away 
(the quasar is just above the top right corner in the first panel of Figure 3).
Note that the vast majority
of the objects without broad-band detection 
appear spatially compact.

In Figure \ref{Stack}, we show a stacked NB (left-hand
panel) and V-band (right-hand panel) image of the 12 proto-galactic
candidates, i.e. the continuum-undetected sources with EW$_0>240\mathrm{\AA}$.
Before stacking, bright sources have been masked everywhere except in the
central 3''-wide region (indicated by the black circle) and the 
local (30'' radius), sigma-clipped mean has been subtracted from each
image.
We then performed aperture photometry within a 2'' diameter aperture 
on both the NB and V-band stack, centered on the position derived from 
the NB. The measured flux in V-band is below the 1$\sigma$ level 
for a 2'' aperture, i.e. $V_{\mathrm{stack}}\sim30.3$, implying 
a combined constraint of EW$_{0,\mathrm{stack}}>800\mathrm{\AA}$ (1$\sigma$).
This measurement strongly suggests that internal stellar sources, if present at all,
do not contribute significantly to the Ly$\alpha$
emission\begin{footnote}{As a reference, we have estimated with
    CLOUDY (v10.00; Ferland et al. 1998) that
    an EW$_0 > 800\mathrm{\AA}$ is still one order of magnitude lower than the value
    expected from pure nebular continuum if measured at rest-frame
    $1600\mathrm{\AA}$ and using $\beta_{\lambda}=-2$ for a fully
    ionized cloud by a typical quasar spectrum ($\alpha=-1.7$;
    e.g. Telfer et al. 2002).}\end{footnote}.
The stacked Ly$\alpha$ emission profile remains compact and consistent with
a point source with FWHM size of about 1'' 
(about 8 physical kpc at $z=2.4$).
 
\begin{figure*}
\begin{center}
\includegraphics[totalheight=0.4\textheight]{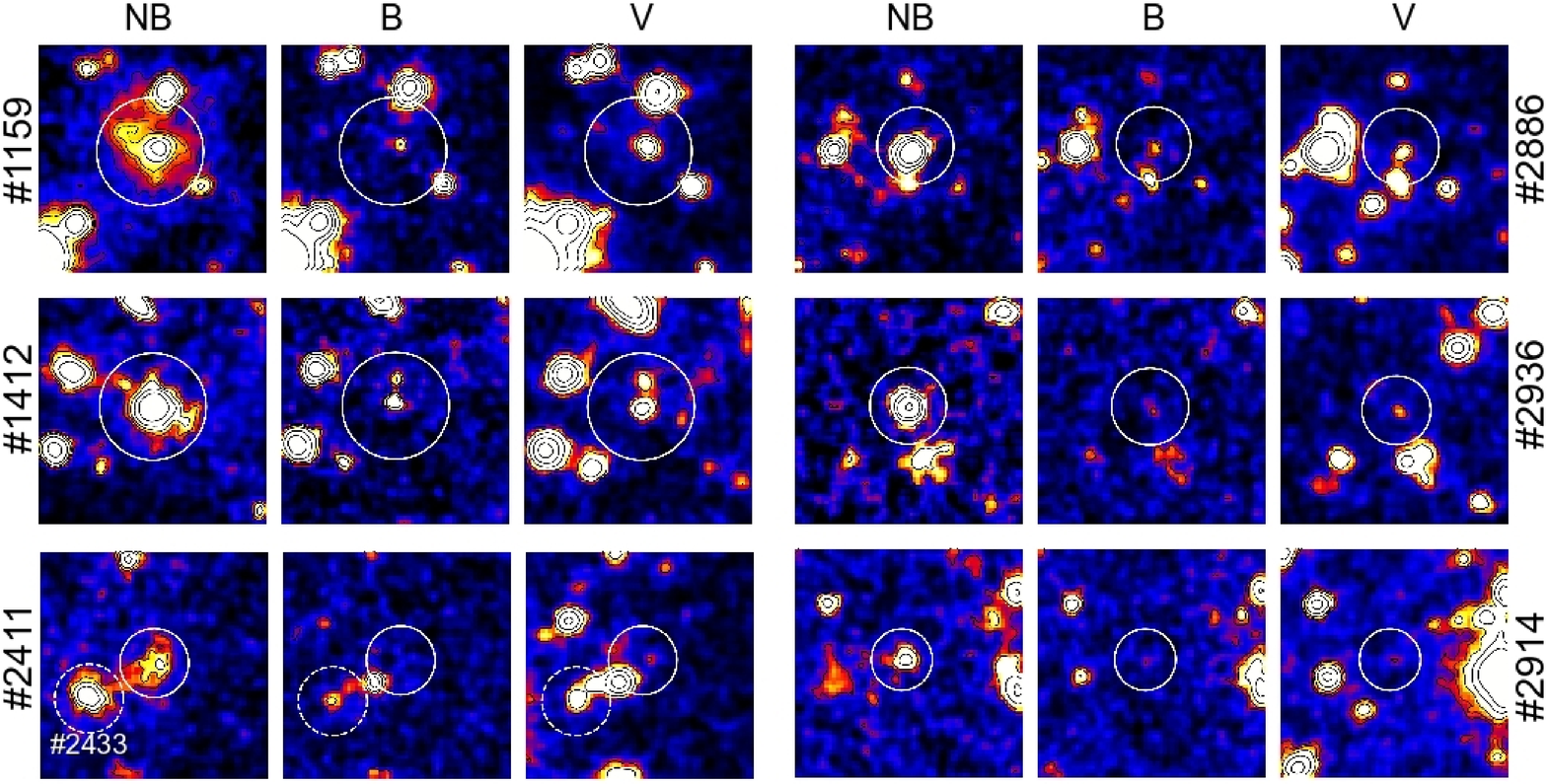}
\caption{Images of the objects with continuum
detection ($V>3\sigma$) but still with EW$_0>240 \angs\ $ (assuming the line is Ly$alpha$)
in three different bands. 
These sources include the extended objects with filamentary emission (\#1159, \#1412, and \#2411 with its ``companion'' \#2433). The high value of the
EW$_0$ suggests that local star-formation cannot be the only source of the detected radiation. Instead, the emission is fully compatible with 
dense circum-galactic gas fluorescently illuminated by the quasar (see
text for details). Panel dimensions are 15 $\times$ 15 arcsec$^2$.   The images have been
smoothed with a gaussian kernel with radius $0".5$ (two pixels) for
clarity.}
\label{LAEFig}
\end{center}
\end{figure*}

Postage stamp images and summary properties 
of the Ly$\alpha$ candidates
with broad-band detections ($>3\sigma$) and measured
EW$_0>240\angs \ $ are presented in Figure \ref{LAEFig}
and Table \ref{LAETable}, respectively.
It can be seen that the three brightest Ly$\alpha$ candidates in our sample, 
\#1159, \#2886 and \#1412
are also the objects with the highest measured
EW, i.e. EW$_0\sim350\angs $. They are detected
both in V and B bands and show a very
red $V-B$ color. In particular, their B magnitudes
are faint enough that the very bright Ly$\alpha$ emission
could contribute most of the broad-band flux. 
The presence of a stellar continuum is favored by the 
clearly detected flux in V-band, but either the stellar
population is old or it is embedded in a large amount of dust
that is not affecting the Ly$\alpha$ emission. This
would be necessary to reconcile the red color with the high
Ly$\alpha$ EW. As we will discuss in section
\ref{CGMsec}, a possible explanation is that
quasar fluorescence is illuminating a large, dust-poor 
gas reservoir - the Circum Galactic Medium (CGM) - 
surrounding dust-rich, star-forming galaxies. 

At least three objects, \#1159, \#1412 and \#2411,
show evidence of extended, possibly 
filamentary, emission. In particular, the image of
\#1159 shows a hint of three 
structures extending out of the bright nucleus. The longest
of these has a (projected) linear size of about 4'', corresponding
to about 30 physical kpc at $z=2.4$.  
This scale is similar to the virial radius of a $\sim10^{11} M_{\odot}$ 
dark-matter halo at this redshift.  Object \#2411 also shows a 
filamentary ``bridge'' on similar (projected) scales apparently 
connecting one region of extended emission without a compact nucleus
(and with only a marginal detection in V-band) to another
bright narrow-band excess object with high EW (\#2433 ).
Revealing the nature of these two apparently connected Ly$\alpha$ 
companions is complicated by the fact that the broad-band
images show a continuum source between them. Without knowing
the redshift and the spectral energy distribution of this continuum
object, it is difficult
to estimate its
contribution to the NB observed flux.
However, the red $B-V$ color and the different morphology 
argue that such
compact object cannot be the only source of narrow-band
emission in the ``bridge''.
If \#2411 and \#2433 are physically connected, this emitting
``bridge'' may have a similar origin to the narrow gas filament 
discovered by Rauch
et al. (2011) in a deep blind spectroscopic survey at $z=3.3$
and corresponding to a continuum source showing tidal tails in
HST imaging. 
 
\begin{figure*}
\begin{center}
\includegraphics[totalheight=0.12\textheight]{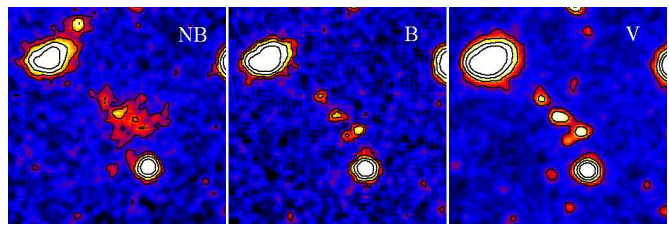}
\caption{Postage stamp images of the extended, blob-like emission LAB1 in narrow-band (left-hand panel), B-band (central panel) and V-band (right-hand panel). 
Panel dimensions are 20$\times$20 arc sec$^2$.   The images have been
smoothed with a gaussian kernel with radius $0".5$ (two pixels) for
clarity. The compact narrow-band excess
object on the top-left corner of the NB image is \#2620, one of the brightest 
fluorescent protogalactic or ``dark'' cloud candidates (EW$_0 > 276\angs\ $) and
undetected in both B and V-band).}
\label{LAB}
\end{center}
\end{figure*}

In Figure \ref{LAB}, we show the narrow and broad-band postage stamp
images of the blob-like extended emitter LAB1.
Morphologically, this object resembles other elongated Ly$\alpha$ blobs
previously discovered (e.g., Steidel et al. 2000, Matsuda et al. 2004,
Nilsson et al. 2006, Yang et al. 2009, Prescott et al. 2012). 
It shows a relatively uniform surface brightness except for
one peak, which is off centered with respect to the extended emission.
At least four continuum sources are detected in the broad-band image(s)
within the projected area of LAB1. One of these sources is located in the proximity of
the NB emission peak. The overall shape of LAB1 seems to follow
these continuum objects but at present we are not able
to confirm if they lie at the same redshift as LAB1.
Note that the bright NB excess object in the top-left corner
in Figure \ref{LAB} (about 10'' away from LAB1) 
is \#2620, one of the objects without continuum-detection with 
high EW limit (EW$_0>276 \angs \ $).

\begin{center}
\begin{table*}
\begin{tabular}{*{10}{c}}
\multicolumn{10}{c}{\textbf{Table 2}: Position and properties of
  Ly$\alpha$ candidates without continuum detection ($V$ and $B<3\sigma$) } \\
\hline
\hline
\multicolumn{1}{c}{Id} &
\multicolumn{1}{c}{RA} &
\multicolumn{1}{c}{DEC}&
\multicolumn{1}{c}{NB$^a$} &
\multicolumn{1}{c}{V$_{lim}^b$}  &
\multicolumn{1}{c}{SNR(V)$^c$}  &
\multicolumn{1}{c}{B$_{lim}^b$}  &
\multicolumn{1}{c}{F$_{Ly\alpha}^d$} &
\multicolumn{1}{c}{L$_{Ly\alpha}^e$} &
\multicolumn{1}{c}{EW$_{0}^f$}
 \\
\multicolumn{1}{c}{ } &
\multicolumn{1}{c}{(J2000)} &
\multicolumn{1}{c}{(J2000)}&
\multicolumn{1}{c}{(AB)} &
\multicolumn{1}{c}{(AB)}  &
\multicolumn{1}{c}{}  &
\multicolumn{1}{c}{(AB)}  &
\multicolumn{1}{c}{10$^{-17}$ erg s$^{-1}$ cm$^{-2}$} &
\multicolumn{1}{c}{10$^{42}$ erg s$^{-1}$} &
\multicolumn{1}{c}{$\mathrm{\AA}$}
\\
\hline
        1484  & 1:11:44.83 & -35:03:12.1 & 25.31$^{+ 0.09}_{-0.08}$ & $>$29.71
  & -1.1& $>$28.16 &  1.95$\pm$ 0.16 &  0.91$\pm$ 0.07 & $>$483 \\
         902  & 1:11:41.80 & -35:04:43.5 & 25.57$^{+ 0.15}_{-0.14}$ & $>$29.75
  & -0.8& $>$29.45 &  1.54$\pm$ 0.20 &  0.71$\pm$ 0.09 & $>$443 \\
        1737  & 1:11:34.45 & -34:59:41.9 & 25.08$^{+ 0.08}_{-0.08}$ & $>$29.22
  &  0.5& $>$29.38 &  2.42$\pm$ 0.17 &  1.12$\pm$ 0.08 & $>$325 \\
         390  & 1:11:43.50 & -35:05:48.7 & 25.67$^{+ 0.14}_{-0.13}$ & $>$29.75
  & -1.5& $>$29.41 &  1.40$\pm$ 0.17 &  0.65$\pm$ 0.08 & $>$315 \\
        2944  & 1:11:45.28 & -35:02:14.1 & 25.42$^{+ 0.11}_{-0.10}$ & $>$29.69
  & -0.3& $>$29.38 &  1.77$\pm$ 0.17 &  0.82$\pm$ 0.08 & $>$303 \\
        1321  & 1:11:51.76 & -35:03:42.0 & 25.87$^{+ 0.17}_{-0.15}$ & $>$29.78
  & -0.3& $>$29.15 &  1.17$\pm$ 0.17 &  0.54$\pm$ 0.08 & $>$297 \\
         555  & 1:11:42.91 & -35:05:24.5 & 25.74$^{+ 0.12}_{-0.11}$ & $>$29.80
  & -0.6& $>$29.47 &  1.32$\pm$ 0.14 &  0.61$\pm$ 0.06 & $>$282 \\
        2620  & 1:11:57.01 & -35:01:35.9 & 25.00$^{+ 0.07}_{-0.07}$ & $>$29.12
  &  0.8& $>$28.62 &  2.60$\pm$ 0.16 &  1.21$\pm$ 0.08 & $>$276 \\
        2131  & 1:11:37.91 & -35:00:27.9 & 25.58$^{+ 0.11}_{-0.10}$ & $>$29.51
  &  0.3& $>$29.54 &  1.52$\pm$ 0.15 &  0.71$\pm$ 0.07 & $>$264 \\
        2274  & 1:11:29.17 & -35:00:47.3 & 26.24$^{+ 0.16}_{-0.14}$ & $>$29.86
  & -1.8& $>$29.54 &  0.83$\pm$ 0.11 &  0.39$\pm$ 0.05 & $>$263 \\
        1519  & 1:11:36.36 & -35:03:17.0 & 25.72$^{+ 0.15}_{-0.14}$ & $>$29.62
  & -0.9& $>$29.31 &  1.34$\pm$ 0.17 &  0.62$\pm$ 0.08 & $>$254 \\
        1212  & 1:11:45.14 & -35:03:57.1 & 26.00$^{+ 0.10}_{-0.09}$ & $>$29.49
  &  0.3& $>$27.91 &  1.04$\pm$ 0.09 &  0.48$\pm$ 0.04 & $>$240
  \\
\hline
        1498  & 1:11:47.81 & -35:03:13.8 & 26.07$^{+ 0.17}_{-0.16}$ & $>$29.71
  & -1.6& $>$29.43 &  0.97$\pm$ 0.14 &  0.45$\pm$ 0.07 & $>$218 \\
        2842  & 1:11:49.17 & -35:02:34.0 & 26.26$^{+ 0.17}_{-0.15}$ & $>$29.68
  & -1.5& $>$29.39 &  0.81$\pm$ 0.12 &  0.38$\pm$ 0.05 & $>$216 \\
        1829  & 1:11:30.15 & -34:59:51.5 & 25.98$^{+ 0.15}_{-0.14}$ & $>$29.80
  & -0.1& $>$28.50 &  1.05$\pm$ 0.14 &  0.49$\pm$ 0.06 & $>$202 \\
        2985  & 1:11:46.39 & -35:02:21.5 & 25.99$^{+ 0.14}_{-0.13}$ & $>$29.69
  & -0.1& $>$29.38 &  1.04$\pm$ 0.13 &  0.49$\pm$ 0.06 & $>$174 \\
        1565  & 1:11:46.79 & -35:03:22.1 & 26.19$^{+ 0.17}_{-0.16}$ & $>$29.71
  & -3.0& $>$29.44 &  0.87$\pm$ 0.13 &  0.40$\pm$ 0.06 & $>$174 \\
        1926  & 1:11:33.06 & -35:00:01.2 & 25.87$^{+ 0.13}_{-0.12}$ & $>$29.47
  &  0.4& $>$28.38 &  1.17$\pm$ 0.13 &  0.54$\pm$ 0.06 & $>$163 \\
         619  & 1:11:58.87 & -35:05:16.2 & 26.04$^{+ 0.15}_{-0.14}$ & $>$29.26
  &  0.6& $>$29.54 &  1.00$\pm$ 0.13 &  0.46$\pm$ 0.06 & $>$156 \\
        2034  & 1:11:43.31 & -35:00:21.3 & 26.13$^{+ 0.15}_{-0.14}$ & $>$29.57
  &  0.2& $>$29.52 &  0.92$\pm$ 0.12 &  0.43$\pm$ 0.06 & $>$146 \\
        1706  & 1:11:56.37 & -34:59:30.7 & 25.92$^{+ 0.16}_{-0.14}$ & $>$29.39
  & -0.2& $>$28.50 &  1.11$\pm$ 0.15 &  0.52$\pm$ 0.07 & $>$140 \\
          85  & 1:11:38.64 & -35:06:27.6 & 26.78$^{+ 0.22}_{-0.19}$ & $>$29.59
  & -1.0& $>$29.04 &  0.50$\pm$ 0.09 &  0.23$\pm$ 0.04 & $>$137 \\
        1279  & 1:11:40.40 & -35:03:45.7 & 24.96$^{+ 0.09}_{-0.09}$ & $>$28.26
  &  2.7& $>$29.39 &  2.70$\pm$ 0.21 &  1.25$\pm$ 0.10 & $>$120 \\
        1439  & 1:11:52.60 & -35:03:05.2 & 26.04$^{+ 0.13}_{-0.13}$ & $>$29.78
  & -0.3& $>$29.06 &  1.00$\pm$ 0.11 &  0.46$\pm$ 0.05 & $>$115 \\
        2454  & 1:11:59.18 & -35:01:11.8 & 26.62$^{+ 0.22}_{-0.19}$ & $>$29.52
  & -0.2& $>$28.62 &  0.58$\pm$ 0.11 &  0.27$\pm$ 0.05 & $>$111 \\
          70  & 1:11:45.10 & -35:06:28.1 & 26.57$^{+ 0.25}_{-0.21}$ & $>$28.58
  &  1.0& $>$28.93 &  0.61$\pm$ 0.13 &  0.28$\pm$ 0.06 & $>$43 \\
        1432  & 1:11:47.69 & -35:03:04.5 & 26.40$^{+ 0.23}_{-0.20}$ & $>$28.21
  &  2.9& $>$29.43 &  0.72$\pm$ 0.14 &  0.33$\pm$ 0.06 & $>$42 \\
        1762  & 1:11:53.86 & -34:59:39.6 & 26.53$^{+ 0.21}_{-0.19}$ & $>$28.53
  &  1.6& $>$29.27 &  0.64$\pm$ 0.11 &  0.30$\pm$ 0.05 & $>$41 \\
        2525  & 1:11:36.41 & -35:01:21.3 & 26.50$^{+ 0.20}_{-0.18}$ & $>$28.41
  &  2.7& $>$29.50 &  0.65$\pm$ 0.11 &  0.30$\pm$ 0.05 & $>$32 \\
         665  & 1:11:51.87 & -35:05:09.6 & 26.56$^{+ 0.19}_{-0.17}$ & $>$28.37
  &  2.9& $>$29.53 &  0.62$\pm$ 0.10 &  0.29$\pm$ 0.05 & $>$32 \\
        1989  & 1:11:38.01 & -35:00:25.4 & 26.47$^{+ 0.31}_{-0.25}$ & $>$27.82
  &  2.8& $>$28.40 &  0.67$\pm$ 0.17 &  0.31$\pm$ 0.08 & $>$28 \\
\hline
\multicolumn{9}{l}{\small{$^a$} Total magnitude.}\\
\multicolumn{9}{l}{\small{$^b$} Lower limits - measured value plus
    the local 1$\sigma$ or local 1$\sigma$ if SNR$<0$ - derived from a circular
    aperture of 1''.5 diameter.} \\
\multicolumn{9}{l}{\small{$^c$} V-band Signal to Noise Ratio within a circular
    aperture of 1''.5 diameter.} \\
\multicolumn{9}{l}{\small{$^d$} Derived from total NB magnitudes
  assuming that emission line is at the filter central wavelength
  (lower limits).} \\
\multicolumn{9}{l}{\small{$^e$} Derived from F$_{Ly\alpha}$ assuming z$=2.4$.} \\
\multicolumn{9}{l}{\small{$^f$} Rest frame EW limits within a 1''.5
  diameter circular aperture estimated from $V_{lim}$ assuming
  an UV continuum slope $\beta_{\lambda}=-2$. } \\
\end{tabular}
\label{PGCTable}
\end{table*}
\end{center}

\begin{center}
\begin{table*}
\begin{tabular}{*{9}{c}}
\multicolumn{9}{c}{\textbf{Table 3}: Position and properties of Ly$\alpha$ candidates with continuum detection ($V>3\sigma$) and EW$_0> 240 \mathrm{\AA}$ $^a$} \\
\hline
\hline
\multicolumn{1}{c}{Id} &
\multicolumn{1}{c}{RA} &
\multicolumn{1}{c}{DEC}&
\multicolumn{1}{c}{NB} &
\multicolumn{1}{c}{V$^b$}  &
\multicolumn{1}{c}{B$^b$}  &
\multicolumn{1}{c}{F$_{Ly\alpha}^c$} &
\multicolumn{1}{c}{L$_{Ly\alpha}^d$} &
\multicolumn{1}{c}{EW$_{0}^e$}
 \\
\multicolumn{1}{c}{ } &
\multicolumn{1}{c}{(J2000)} &
\multicolumn{1}{c}{(J2000)}&
\multicolumn{1}{c}{(AB)} &
\multicolumn{1}{c}{(AB)}  &
\multicolumn{1}{c}{(AB)}  &
\multicolumn{1}{c}{10$^{-17}$ erg s$^{-1}$ cm$^{-2}$} &
\multicolumn{1}{c}{10$^{42}$ erg s$^{-1}$} &
\multicolumn{1}{c}{$\mathrm{\AA}$}
\\
\hline
 1159  & 1:11:56.59 & -35:03:46.9 &  22.66$^{ 0.04 }_{ -0.04 }$ &  24.74$^{+ 0.04 }_{ -0.04 }$ &  26.06$^{+ 0.13 }_{ -0.12 }$ & 22.44$\pm$0.81 & 10.42$\pm$0.38 &  358$\pm$35  \\
 2886  & 1:11:46.88 & -35:02:33.8 &  23.18$^{ 0.04 }_{ -0.04 }$ &  25.99$^{+ 0.07 }_{ -0.06 }$ &  26.72$^{+ 0.17 }_{ -0.15 }$ & 13.90$\pm$0.50 &  6.46$\pm$0.23 &  356$\pm$48  \\
 1412  & 1:11:45.42 & -35:03:05.3 &  22.47$^{+0.04 }_{ -0.04 }$ &  25.45$^{+ 0.06 }_{ -0.06 }$ &  25.98 $^{+ 0.11 }_{ -0.11 }$ & 26.73$\pm$0.97 & 12.42$\pm$0.45 &  327$\pm$28  \\
 2411  & 1:11:33.48 & -35:01:09.8 &  24.25$^{+ 0.06 }_{ -0.06 }$ &  27.81$^{+ 0.22 }_{ -0.18 }$ & $>$ 28.73  &  5.19$\pm$0.28 &  2.41$\pm$0.13 &  296$\pm$51  \\
 2914  & 1:11:33.93 & -35:02:13.0 &  25.14$^{+ 0.08 }_{ -0.08 }$ &  28.69$^{+ 0.34 }_{ -0.27 }$ & $>$ 28.52  &  2.29$\pm$0.16 &  1.06$\pm$0.08 &  291$\pm$77  \\
 2936  & 1:11:53.84 & -35:02:18.1 &  23.90$^{+ 0.05 }_{ -0.05 }$ &  27.34$^{+ 0.24 }_{ -0.2 }$ & $>$ 27.98  &  7.16$\pm$0.32 &  3.33$\pm$0.15 &  262$\pm$48  \\
\hline
 2433$^f$  & 1:11:33.85 & -35:01:12.4 &  23.32$^{+ 0.09 }_{ -0.08 }$ &  25.35$^{+ 0.12 }_{ -0.11 }$ &  26.3 $^{+ 0.11 }_{ -0.1 }$ & 12.22$\pm$0.97 &  5.68$\pm$0.45 &  218$\pm$27  \\
\hline
\multicolumn{9}{l}{\small{$^a$ See Appendix A for the list of the remaining
   61  Ly$\alpha$ candidates with continuum detection.}} \\
\multicolumn{9}{l}{\small{$^b$ Total magnitudes (or lower limits
    for B-band, see Table 2) .}} \\
\multicolumn{9}{l}{\small{$^c$} Assuming that emission line is at the filter central wavelength.} \\
\multicolumn{9}{l}{\small{$^d$} Derived from F$_{Ly\alpha}$ assuming z$=2.4$.} \\
\multicolumn{9}{l}{\small{$^e$} Rest frame EW for Ly$\alpha$ at z$=2.4$. A UV continuum slope $\beta_{\lambda}=-2$ is used for objects undetected in B. } \\
\multicolumn{9}{l}{\small{$^f$} (Projected) ``Companion'' of \#2411. } \\
\end{tabular}
\label{LAETable}
\end{table*}
\end{center}

\begin{figure*}
\begin{center}
\includegraphics[totalheight=0.3\textheight]{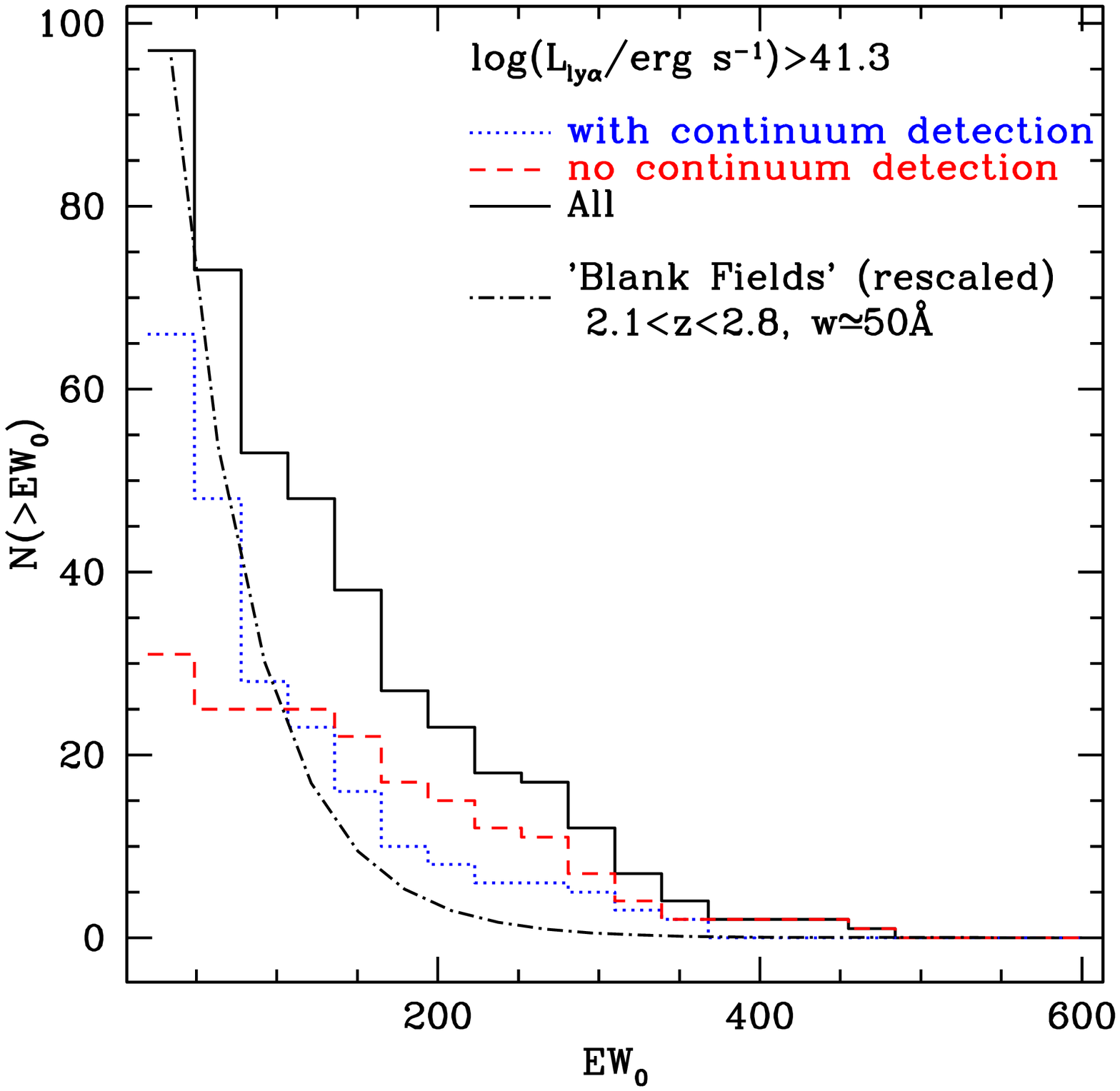}
\includegraphics[totalheight=0.3\textheight]{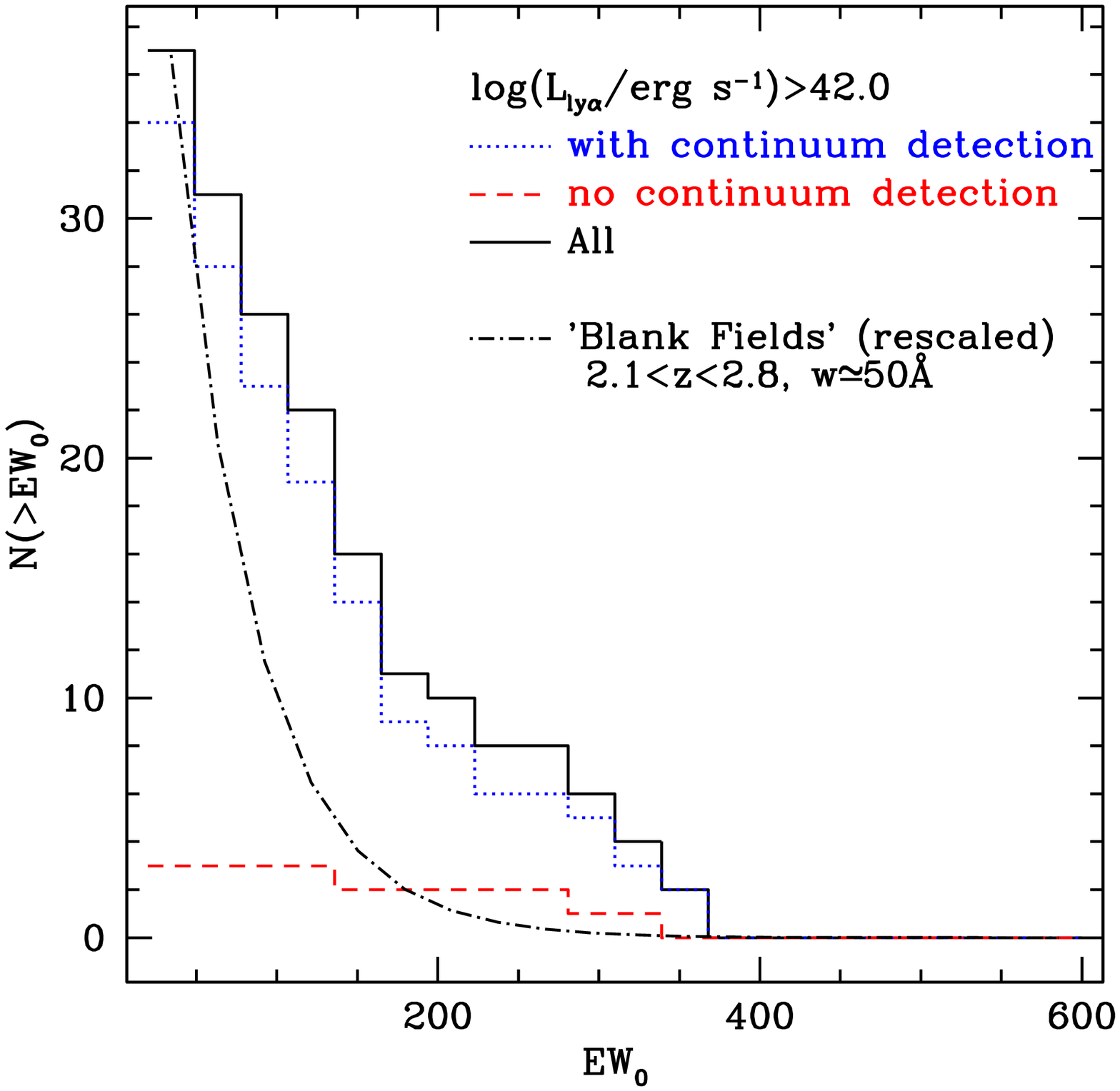}
\caption{The cumulative rest frame equivalent width (EW$_0$) distribution
for all the detected line emitting objects in our sample (left-hand panel; $L > 2\times
10^{41}$ erg s$^{-1}$) and considering only the candidates with
$L_{Ly\alpha} > 10^{42}$ erg s$^{-1}$ (right-hand panel), always assuming the line is Ly$\alpha$.
Objects are split according to their detection (blue dotted
histogram) or non-detection (red dashed histogram) in any of the
broad-bands. Note that the EW$_0$ of the objects without continuum
detections are lower limits. The solid black histogram shows the
cumulative EW distribution including all objects and using the lower
limits for the continuum-undetected candidates.
The results from recent surveys 
targeting 'blank-fields', 
at $2.1<z<2.8$ are shown (black dot-dashed line, rescaled to match
the total number of objects) taken from Ciardullo et al. 2012 ($z=2.1$,
$L_{Ly\alpha}>10^{42}$ erg s$^{-1}$;
original data from Guaita et al. 2010), Nilsson et al. 2009 ($z=2.3$, 
$L_{Ly\alpha}>2\times 10^{42}$ erg s$^{-1}$), and
Grove et al. 2009 ($z=2.8$, $L_{Ly\alpha} > 3\times 10^{41}$ erg
s$^{-1}$; 
original data from Fynbo et al. 2003).
The observed EW distributions from these surveys can all 
be fitted by a single exponential function with an e-folding scale
length of $w\simeq50\angs\ $ (see text for details).
Our sample presents a clear excess of high EW objects with respect to
the 'blank-fields' surveys. This is true also considering only the
continuum-detected objects 
or the more luminous part of the sample (right hand panel),
This clear excess of high EW$_0$ with respect to the field surveys is a distinctive sign 
of a fluorescent Ly$\alpha$ boosting due to the quasar proximity. 
}
\label{EWdist}
\end{center}
\end{figure*}

\section{Physical origin of the emission}\label{originsec}

In this section, 
we compare our observational 
results to those obtained in similar
deep searches for Ly$\alpha$
emission in regions that do not
target high-redshift quasars
(we refer to these surveys as ``blank-field''
studies) and with the results  
of radiative-transfer simulations. 
In particular, we will examine the
EW distributions, Luminosity Functions
and the relation between (projected) distances
form the quasar and candidate luminosities.
The combination of these statistics
provide strong evidence 
for the fluorescent origin of the line emission of many of the objects in
our sample.

\subsection{EW distribution}

The Ly$\alpha$ EW is one of the most
important diagnostics to
distinguish between internal star-formation
and other mechanisms that can produce
Ly$\alpha$ emission (e.g., Schaerer 2002), simply because of the limit
to the ratio of the number of ionizing photons to continuum photons 
at the Ly$\alpha$ wavelength from possible stellar populations.
In the proximity of a quasar: the Ly$\alpha$ emission may
be boosted by fluorescent emission,
leaving untouched the stellar continuum,
if present.
The net effect is an EW ``boost''
of every object directly illuminated
by the quasar by an amount that
may vary with the object gas mass
and its distance from the quasar.
Therefore, if a significant fraction of
our objects are indeed fluorescently boosted,
our EW distribution should be
skewed towards higher values with 
respect to surveys away from bright
quasars.

Slightly different observational methods and
selection techniques complicate 
the comparison to surveys
in the literature.
Despite these difficulties, 
the observed EW distributions from several recent 'blank-fields' surveys 
at $2.1<z<2.8$ are remarkably similar.
In particular, they
can all be fitted by a single exponential function with an e-folding scale
length of $w\simeq50\angs\ $. 
These surveys include:
Grove et al. 2009 ($z=2.8$, $L_{ly\alpha}>3\times 10^{41}$ erg
s$^{-1}$; 
original data from Fynbo et al. 2003\begin{footnote}
{ Extracted from a field containing a damped Ly$\alpha$ system 
at $z=2.85$ (Q$2138-4427$); we computed the EW distribution 
using all photometrically selected candidates in this field
(Tables A3 and A4 in Grove et al. 2009).
Note that,
although this is not a strictly blank field since it contains
a DLA, the over density of Ly$\alpha$ candidates 
is only a factor 1.1 with respect to the
other surveys.
}\end{footnote}
)
, Nilsson et al. 2009 ($z=2.3$, 
$L_{ly\alpha}>2\times 10^{42}$ erg s$^{-1}$), and
Ciardullo et al. 2012 ($z=2.1$,
$L_{ly\alpha}>10^{42}$ erg s$^{-1}$;
original data from Guaita et al. 2010)
\begin{footnote}{ The only
exception is given by the results of Hayes et al. ($z=2.2$,
$L_{ly\alpha}>3\times 10^{41}$)
who find a slightly larger e-folding scale with value 
$w\simeq76\angs \ $ but 
they suggest that poor statistics are likely
responsible for this discrepancy.}
\end{footnote}
. 
As shown by these studies, the EW
distribution seems to be also largely 
independent of the luminosity cut
except at the very bright end, where there is 
an apparent lack of high EW objects.
The EW distribution appears instead to evolve with redshift:
at $z\sim3.1$ there are evidences
for an increase in EW scale length to $w\simeq70\angs$,
(Ciardullo et al. 2012; Nilsson et al 2009; see also Ouchi et al. 2008) 
or larger at $z>4$
(e.g., Malhotra \& Rhoads 2002, Dawson et al. 2004, Saito et al. 2006, Shimasaku et al. 2006).

In Figure \ref{EWdist}, we present the cumulative EW distribution 
for all the detected objects in our sample (left-hand panel; $L>2\times
10^{41}$ erg s$^{-1}$) and considering only the candidates with
$L_{ly\alpha}>10^{42}$ erg s$^{-1}$ (right-hand panel).
The sample is divided according to whether the continuum was detected (blue dotted histogram) 
or not (red dashed histogram). 
The EW$_0$ of the objects without continuum
detections are of course therefore lower limits. The solid black histogram shows the
cumulative EW distribution including all objects and using the lower
limits for the continuum-undetected candidates.

Our sample contains a clear excess of high EW objects with respect to
the 'blank-field' surveys. 
This remains true also considering only the
continuum-detected objects (that show an e-folding scale of 
$w\simeq85\angs\ $) or
when we consider 
only the most luminous part of the sample:
in this case the e-folding scale
for the continuum-detected objects (the large majority) is of the
order of $w\simeq120\angs $, although the statistics are poorer.
In both cases,
a K-S test confirms that our EW distribution is 
different from the one obtained by the ``blank-field'' surveys
at a significance greater than 99\% . 
It should be noted that we that we have not attempted to correct for the non-square bandpass
of our filter since we have assumed that
the Ly$\alpha$ emission of each candidate lies at the
peak transmission wavelength of the filter (since we do not know
the redshift distribution for all of our objects). 
We stress that this is a conservative assumption for the calculation of the
EW and likely gives an underestimate of the true Ly$\alpha$
flux and EW of our candidates. 

In addition to explaining the excess of high EW objects, 
the fluorescent Ly$\alpha$ boosting due to the proximity of the bright quasar
is also able to account for the $w$ increase with luminosity
that is exhibited by those objects with detected continua, the opposite trend to that 
exhibited by the ``blank-field'' surveys.
Quasar fluorescence boosts \emph{both} the
Ly$\alpha$ luminosity and the EW 
of a given object (independent of its ``original'' 
position in the L$_{ly\alpha}$-EW diagram )
creating a positive correlation between these two quantities.

Is there any other effect that could 
explain the different EW distribution
obtained in our sample with respect
to the ``blank-field'' surveys? 
Differently from the ``blank-fields'', our
sample could be associated with a 
high density peak in the matter
distribution due to the presence
of the quasar. We discuss extensively
this possibility in the following sections. 
Venemans et al. 2007 performed 
a survey for Ly$\alpha$ emitters around
high-redshift radio galaxies detecting
a significant enhancement with respect
to the field (about a factor 2-5). 
However, their EW distributions do not
appear significantly different
from the one obtained by the
``blank-field'' surveys.
In particular, the fraction of candidates
with EW$_0>240\angs \ $ (and $L_{ly{\alpha}}>10^{42}$
erg s$^{-1}$) is less than 5\% (B. Venemans,
private communication), similarly to
the ``blank-field'' surveys 
at $2<z<3$. As a reference, 
we find instead that more than 20\% of our
sources with
$L_{ly\alpha}>10^{42}$ erg s$^{-1}$
have EW$_0>240\angs \ $.

\begin{figure}
\begin{center}
\includegraphics[totalheight=0.3\textheight]{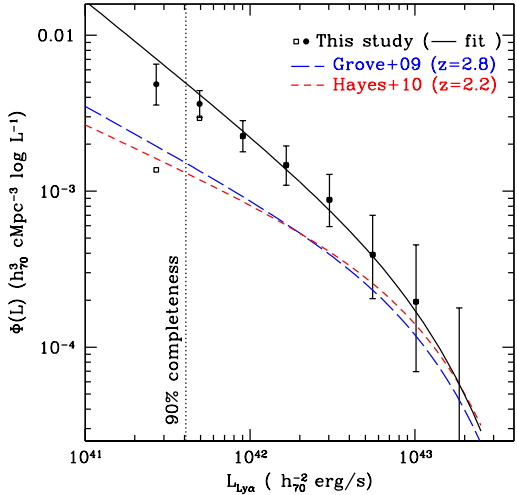}
\caption{
The luminosity function (number density per decade in luminosity) of our
Ly$\alpha$ candidates  compared with other two deep surveys
in ``blank-fields'' (see labels within figure).
Solid circles (or open squares) show the observed Luminosity Function
including (or excluding) the correction for completeness.
The error bars show 84\% confidence levels based on Poisson and binomial
statistics (Gehrels 1986). Note that, consistently with the other
surveys, we have not corrected for the non-square bandpass shape of the NB
filter (see text for details). We detect a clear excess of sources at
the faint end with respect to ``blank-fields''. As discussed in
section 4.2, this is fully compatible with fluorescent boosting due to
quasar proximity.
}
\label{LF}
\end{center}
\end{figure}

\begin{figure}
\begin{center}
\includegraphics[totalheight=0.3\textheight]{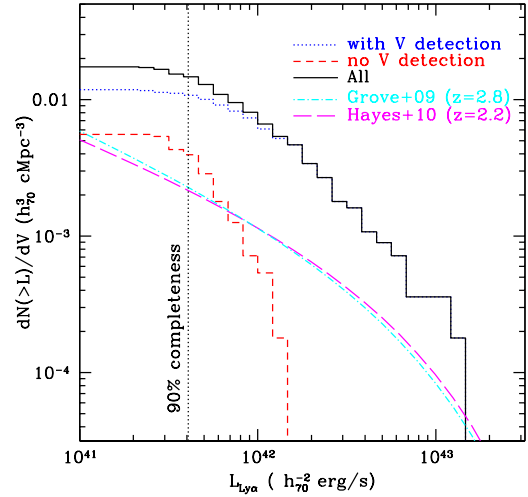}
\caption{Cumulative luminosity function (CLF) of our Ly$\alpha$
candidates (black solid histogram) compared with other deep surveys
in the field (see labels within figure). Blue dotted (or red dashed)
histogram shows the CLF including only objects with (or without) continuum-detection.
}
\label{CLF}
\end{center}
\end{figure}

\begin{figure}
\begin{center}
\includegraphics[totalheight=0.3\textheight]{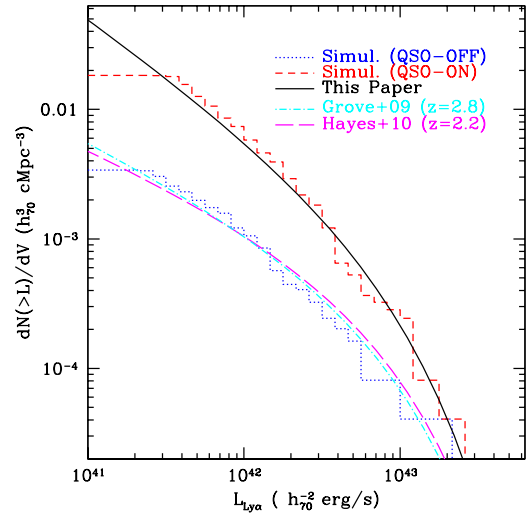}
\caption{Simulated cumulative luminosity function (CLF) obtained from simulations  
assuming that the quasar is off (blue dotted histogram), i.e. no
increase in the UV background due to quasar proximity, and that the quasar is turned on
(red dashed histogram). These simulations include full three-dimensional radiative transfer of
ionizing and Ly$\alpha$ photons (Cantalupo \& Porciani 2011; Cantalupo et al. 2005) 
within high-resolution cosmological simulations of structure formation around a massive
halo (see text for details). The overall vertical normalization of the CLF of Ly$\alpha$ emitters in the QSO-OFF simulation has been rescaled 
to match the observed field CLF of deep surveys (Grove et al. 2009, Hayes et al. 2010), while the slope of the CLF
is reproduced naturally by our simulation. Using the same normalization factor the numerical model is able to reproduce the increase in the total number of
objects and the steepening of the CLF as observed in our data once the quasar is turned on.}
\label{CLFsim}
\end{center}
\end{figure}

\subsection{Luminosity Function}\label{LFsection}

Similarly to the EW distribution, we expect
fluorescent boosting to also impact 
the Ly$\alpha$ Luminosity Function (LF)
of our sample.
In Figure \ref{LF},  we show the LF obtained from
all the Ly$\alpha$ candidates 
(open squares) and including a correction for
completeness (solid circles; see section \ref{LyaSel}). We compare our LF with the
results of Grove et al. 2009 and Hayes et al. 2010.
These two surveys, the deepest available at $2<z<3$, 
reached a comparable depth and used a very similar NB filter on VLT-FORS
with respect to our study.
Despite the small volumes sampled, these two studies derived
very similar luminosity functions, consistent with a Schechter function 
with log$(L^*/$erg s$^{-1}$)$\sim43.1$, 
slope $\alpha\sim-1.5$ and normalization
log$(\phi^*/$cMpc$^{-3})\sim-3.9$.

In our own survey, we detect a clear excess of Ly$\alpha$
candidates with respect to these ``blank-fields''. Fixing
$L^*$ and the normalization to the same value of Grove et al. 2009
and Hayes et al. 2010 our data is best fit by a substantially
steeper faint-end slope ($\alpha\simeq-1.9$). 
As shown by the Cumulative Luminosity Function (CLF) in Figure \ref{CLF},
this excess of Ly$\alpha$ candidates is mostly driven
by objects with continuum detections (blue dotted histogram),
while the CLF  for candidates without continuum detection 
(red dashed histogram) steeply rises only below $L<10^{42}$ erg s$^{-1}$.
The total number density above our 90\% completeness limit
(corresponding to $L_{Ly\alpha}>4\times10^{41}$ erg s$^{-1}$)
is $n\sim0.014$ cMpc$^{-3}$, is about a factor 5 higher than 
Grove et al. 2009 and Hayes et al. 2010 at similar luminosities. 

An enhancement in the number of Ly$\alpha$ emitters (LAE), i.e. in $\phi$* and possibly in L*, may be expected
around a quasar simply because this is likely to occupy a high density region in the Universe. 
For instance, Venemans et al. (2007) found that LAEs around radio galaxies
tend to be of a factor 2-5 more numerous with respect 
to the ``field''.  A similar effect may be playing a role in our sample,
although we note that radio-quiet
quasars (like \QSOtwo\ ) do not generally seem to 
show a different environment with respect to
``normal'', $L^*$  galaxies, at least on
scales larger than 0.5 cMpc (e.g., Croom \& Shanks 1999,
Serber et al. 2006, Falder et al. 2011; see also Hennawi et al. 2006).
A very recent survey by Trainor \& Steidel (2012),
obtained the same result also for hyper-luminous quasars (like
\QSOtwo\ ) at
$z\sim2.7$, concluding that the very low space density of such
objects results from an extremely rare event on scales $<<1$ cMpc and
not from being hosted by rare dark matter haloes. 

Nevertheless, we also
observe a steepening of the LF.  
As previously discussed, 
fluorescent boosting should enhance 
the Ly$\alpha$ emission 
of a gas rich object independent
of its SFR (or any other combination of parameters
that determines the Ly$\alpha$ luminosity
of an isolated galaxy). This constant additive boost to the Ly$\alpha$ flux will have a larger relative effect at low luminosities, thereby steepening the faint end slope $\alpha$.  This steepening is the clearest signature of fluorescent emission in the luminosity function. 

In order to understand if the observed 
steepening of the LF is 
compatible with fluorescence,
we performed a numerical experiment
with the help of cosmological
hydrodynamical simulations 
(using the hydro-code RAMSES, Teyssier 2002)
combined with full three-dimensional 
radiative transfer for continuum 
(RADAMESH, Cantalupo \& Porciani 2011)
and Ly$\alpha$ 
radiation (Cantalupo et al. 2005; Cantalupo et al. 2007).
The simulations will be discussed
in detail in a companion paper (Cantalupo et al., in preparation).
In Appendix B, we briefly describe the numerical methods and
simulation parameters. 

The simulations have been used to produce ``mock'' observations
of Ly$\alpha$ emitting objects in a $40^3$ cMpc volume 
of the Universe at $z=2.4$ around a quasar. 
The latter has been turned on and off in order to simulate 
the differential effect of fluorescent boosting on the LF.
The results are presented in Fig. \ref{CLFsim}. In the ``QSO-OFF''
simulation we have assigned a Ly$\alpha$ emissivity
to each object proportionally to its SFR (see Appendix B)
plus the recombination radiation due to the UVB background
ionizations, i.e. UVB fluorescence (several orders of magnitude
below our detection threshold).
Then, we applied our Ly$\alpha$ RT, added the observed
noise to the simulated image and selected candidates
using the same SExtractor procedure applied to the real data.
The LF obtained in this way (blue histogram) naturally reproduces the observed
shape of the LF in ``blank-fields'' once a normalization factor
$n_{\mathrm{ly\alpha}}=0.13$ is applied to take into account the fraction of galaxies
appearing as Ly$\alpha$ emitters and the overdensity of our simulation box
(see Appendix B). 
We then turned on a quasar in the most massive halo in the
simulation (located at its center) and propagated its ionizing photons
through hydrogen and helium within the box using our radiative transfer code RADAMESH. 
We then repeated exactly the procedure followed for the
``QSO-OFF'' simulation, using the same parameters and normalization
factor of the latter. Thus the only difference between
the two simulations is the presence of the quasar ionization field. 
The resulting LF in the ``QSO-ON'' case is shown as a red histogram
in Figure \ref{CLFsim}.
The steepening in the simulated LFs is clearly visible and is 
very similar to the difference between the observed LF in 
the ``blank-field'' surveys and in our sample around the quasar.
Within the limitation of numerical simulations and
observational errors, 
this provides an additional support for the idea that fluorescence is playing a major role in our sample

\begin{figure}
\begin{center}
\includegraphics[totalheight=0.3\textheight]{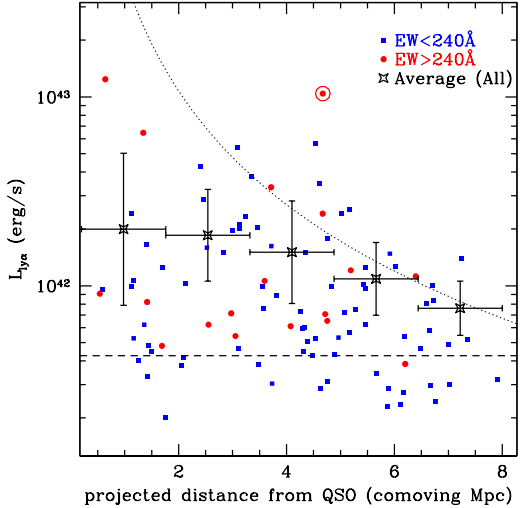}
\caption{The luminosity of Ly$\alpha$ candidates as a function of projected distance from the quasar.
The dotted line shows the expected fluorescent Ly$\alpha$ emission produced by an optically thick cloud of 1 arcsec$^2$
area as a function of its three-dimensional distance from the quasar. This represents an upper limit for
fluorescent protogalactic clouds without star formation at the quasar redshift. The luminosity would scale linearly with object area, but the majority of detected objects are compact. One notable exception is \#1159 (circled),
that has an area of $\sim20$ arcsec$^2$.
The dashed line represents the 90\% completeness limit of the survey. We have used all the objects above this limit (except the
very extended \#1159) to compute the \emph{average} luminosity per projected distance bin (starred open squares).
If a significant number of optically thick clouds are present and the objects have Ly$\alpha$ emission boosted
by quasar fluorescence, then we expect the average luminosity to decrease approximatively 
with the square of the distance from the quasar (assuming no intrinsic relation between Ly$\alpha$ emission from 
star formation and distance from the quasar). Despite the large errorbars, there is a marginal detection of the expected signal. 
This is also confirmed by the results of our radiative transfer simulations
presented in figure \ref{Ldist_sim}.
}
\label{Ldist}
\end{center}
\end{figure}

\begin{figure}
\begin{center}
\includegraphics[totalheight=0.3\textheight]{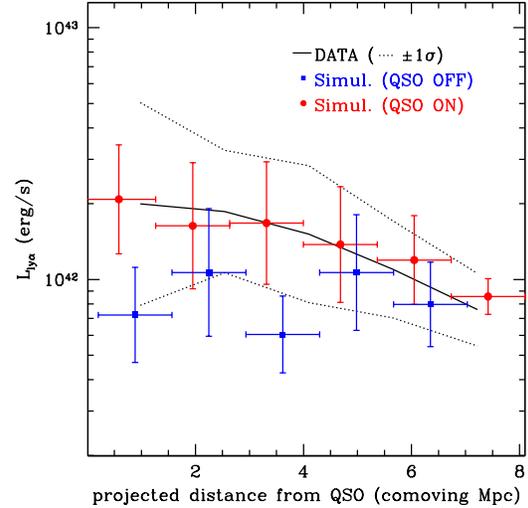}
\caption{Comparison between the average luminosity of observed Ly$\alpha$ candidates (black solid line) and
the average luminosity of the simulated Ly$\alpha$ emitters (blue squares for QSO-OFF and red dots for QSO-ON simulations) 
as a function of projected distance from the quasar above the same
luminosity cut as in figure \ref{Ldist}. Despite large errorbars, the observed trend
is very similar to the QSO-ON simulation suggesting that the observed signal is compatible with
fluorescent emission boosted by the quasar proximity.
}
\label{Ldist_sim}
\end{center}
\end{figure}

\subsection{Luminosity-Distance Relation}

An important prediction for quasar fluorescence
is the relation for optically thick clouds between the fluorescent Surface Brightness (SB)
and the (three-dimensional) distance from the quasar
(e.g., Cantalupo et al. 2005).  This is because the fluorescent boost is proportional to the apparent brightness of the quasar as seen by the cloud.
Optically thick objects whose
size is much larger than the ionized region
act like a ``mirror'' of the impinging
quasar radiation (Gould \& Weinberg 1996).
Their fluorescent SB would thus simply scale
like the inverse square of their three-dimensional
distance from the quasar, assuming isotropic
and temporally constant quasar emission.
In reality, however, geometrical and kinematical effects 
can reduce the expected
SB by a large factor, although this reduction can be parametrized
as shown in Cantalupo et al. (2005).

Furthermore, the SB of those gas clouds that have been completely ionized
by the quasar
will be independent of distance from the quasar (for as long as they are completely ionized).
Their luminosity is simply
proportional to the 
integral of their hydrogen density squared (see section \ref{gasmasssec}).
Finally, any local star formation 
might add a substantial Ly$\alpha$ flux to the 
fluorescent component, and this also would be expected to be independent of distance to the quasar.

Despite these caveats, if a significant fraction of optically
thick clouds (``fluorescent mirrors'') 
are present,
we might still expect
to find an 
increase of the \emph{average}
Ly$\alpha$ SB in our sample closer to the quasar. 
Unfortunately, we cannot obtain directly either the surface brightness
or the three-dimensional distance from the quasar
for most of our objects. This is due to their
compact nature and
the lack, so far, of spectroscopic data for the whole sample
(a spectroscopic follow-up
of the brightest objects will be presented in
Cantalupo et al., in prep.).
However, we can use the total Ly$\alpha$ luminosity and the {\it projected} distance from the quasar as proxies for these two
quantities, 
as shown for our sample in  
Fig. \ref{Ldist}. 
Although the statistics are poor, we do find indeed that
the average luminosity above the completeness level
decreases with the projected distance from 
the quasar (starred open squares).

As a reference, we have
plotted on the same figure 
the expected ``mirror'' fluorescent Ly$\alpha$ luminosity
of a optically thick cloud with an area of
1 arcsec$^2$ as a function
of three-dimensional distance from the quasar
(dotted line). 
This will represent the ``maximum''
emission for a purely fluorescent
source of this area (e.g. Cantalupo et al. 2005).
Although the uncertainties related to
projection effects and object sizes
are quite large, 
it is interesting to note that
the expected trends are similar to those observed, except in the close
proximity of the quasar.
This is the region where
we would expect the quasar radiation to be so intense that
very few clouds would be able
to remain optically thick.

A better analysis that included the size, projection
and ionization effects can be performed
with the help of the radiative transfer simulations
presented in the previous section
and discussed in detail in Appendix B.
In Fig. \ref{Ldist_sim}, we show
the \emph{differential} effect
of fluorescent boosting on the
simulated projected
distance-luminosity relation
once the quasar is turned-on.
The trend obtained from the QSO-ON simulations
(red dots) is now in much better agreement 
with the observations (solid line) at all radii.
Note that this trend was not present
without the effect of the quasar radiation
(blue dots).
Within the limitations of our numerical models,
this is a further
support that
fluorescent boosting is playing a significant role in the Ly$\alpha$ emission of the objects detected in our survey.

\section{The nature of the sources}\label{naturesec}

The statistical analyses performed in the previous 
section indicate that
the dominant emission mechanism 
for many of our sources is likely to be
Ly$\alpha$ fluorescence induced by the quasar.  
This mechanism then gives us a unique opportunity to
directly image, in emission, dense gas
independent of any associated star formation.
This feature distinguishes these new observations 
from previous surveys for
Ly$\alpha$ emitting galaxies or other structures. 
Moreover, 
fluorescent boosting increases the Ly$\alpha$
luminosity of every galaxy, if illuminated by the
quasar that is gas rich. In principle, this makes intrinsically
faint Ly$\alpha$ emitters, that would have been
below our detection threshold, visible in our
survey.  This raises the obvious question: what are the nature and physical properties of this new population of sources
and how do they differ from previously discovered Ly$\alpha$ emitters?

\subsection{Proto-galactic clouds and faint galaxies}

As previously discussed, a high Ly$\alpha$ EW (e.g., EW$_0 > 240\angs \ $ )
is the first criterion to exclude objects in which Ly$\alpha$ emission is
powered (solely) by star-formation. In our sample of 98 narrow-band
excess objects, we detected 18 sources with equivalent width measurements or lower limits above EW$_0>240\mathrm{\AA}$

Among these 18 objects with demonstrated EW$_0 > 240\mathrm{\AA}$, all but six do not have any
detectable continuum counterpart and therefore quasar fluorescence
is likely the dominant source of Ly$\alpha$ photons.  Their Ly$\alpha$ luminosities, which are
all around or below 10$^{42}$ erg s$^{-1}$ (see Figure \ref{CLF}) are
also fully compatible with fluorescence given their distance from
the quasar and their compact sizes, as shown in Figure \ref{Ldist}. 

It should be noted in passing that a further 19 objects do not have any continuum detection 
and could also be entirely powered by quasar fluorescence. However,
the broad-band images 
are not deep enough to require their EW$_0$ to lie above the 240$\angs \ $
level.

It is clear from Figure \ref{CLF} that the steepening of the cumulative Luminosity Function with respect
to the ``blank-surveys'' involves both objects with and without continuum detections - it does not come from the non-continuum detected objects alone, and is mostly due to the larger number of continuum-detected objects
that are also presumably boosted by quasar fluorescence.
The existence of these fluorescently boosted continuum-detected galaxies
is confirmed by the EW distributions presented in 
Figure \ref{EWdist} especially when we restrict to the bright
sample with L$_{ly\alpha}>10^{42}$ erg s$^{-1}$. 
As discussed in section \ref{LFsection}, our radiative
transfer simulations naturally reproduce the steepening
of the cumulative luminosity function 
(see Figure \ref{CLFsim}) and the variation of the average
Ly$\alpha$ luminosity with distance from the quasar (see Figure
\ref{Ldist_sim})
once quasar radiation is ``turned-on''.
It is therefore instructive to have a closer look at the
physical properties of the simulated sources. 

\begin{figure}
\begin{center}
\includegraphics[totalheight=0.3\textheight]{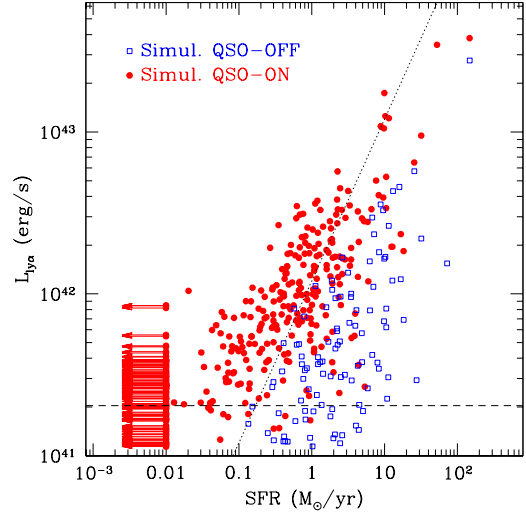}
\caption{Luminosity versus SFR for the Ly$\alpha$ emitters in the QSO-OFF (blue squares) and QSO-ON (red circles) simulations.
In the QSO-OFF simulations, the intrinsic Ly$\alpha$ emissivity is set
proportional to the SFR (
see Appendix B for details). 
After adding simulated noise, the object luminosities are obtained with
SExactor aperture
photometry, similarly to the treatment of the actual data.
Note that the simulations include the full three-dimensional
scattering
of Ly$\alpha$ photons within the circumgalactic and intergalactic
medium (Cantalupo et al. 2005)
but they do not include dust absorption.
Without radiative transfer
effects, the Ly$\alpha$ emitters in the QSO-OFF
simulation would therefore lie on the intrinsic luminosity-SFR relation given by the dotted black
line, which represents a ``maximum'' value for the recovered Ly$\alpha$
emission powered by star formation.
Quasar fluorescence in the QSO-ON simulation (red dots) boosts many objects above this limit,
especially at lower SFRs,
generating the steepening of the CLF observed in Figure \ref{CLFsim}.
About 20\% of sources in the simulation lying above our nominal detection
threshold (dashed line) are not associated with star-formation
within the simulation (at least above our numerical
resolution limit of $\sim 0.01 M_{\odot}$ yr$^{-1}$). They are
represented by left-pointing arrows at SFR$=0.01$.
Their relative abundances and luminosities
are very similar to the observed objects without continuum detection
in our survey strengthening the association
between the detected continuum-less sources 
and proto-galactic clouds. 
}
\label{SFRLum}
\end{center}
\end{figure}

\begin{figure}
\begin{center}
\includegraphics[totalheight=0.3\textheight]{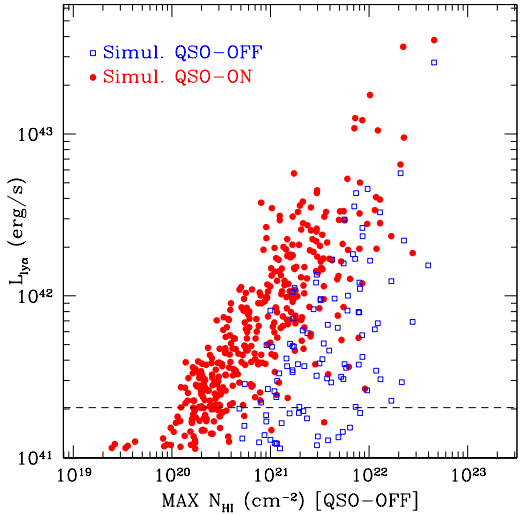}
\caption{
The relation in the simulations between the Ly$\alpha$ luminosity and the maximum
HI column density \emph{before} the quasar is turned on 
(see labels within the panel). The maximum
N$_{\mathrm{HI}}$ is defined considering all the 
lines-of-sight that intersect any image pixel with detectable
Ly$\alpha$ emission. Note that we do not simulate
the formation of H$_2$, and thus the reported HI
values will be an overestimate, especially 
at N$_{\mathrm{HI}}>10^{21}$ cm$^{-2}$.  
}
\label{NHILum}
\end{center}
\end{figure}

In Figure \ref{SFRLum} we show the relation between the
SFR and Ly$\alpha$ luminosity of the simulated
emitters. 
This has been obtained performing SExtractor aperture photometry 
(after adding artificial
noise to the image) in the same way as with the actual data.
In the ``QSO-OFF'' simulation, the only
``source'' of Ly$\alpha$ photons are the stars. The intrinsic Ly$\alpha$ emissivity
is set proportional to the SFR (see appendix B for
details).  In the absence of radiative transfer effects the
simulated objects in the ``QSO-OFF'' simulation
(blue open squares)
would thus lie on the relation indicated by the black dotted line.
It should be noted that our simulations do not include absorption by
dust particles, and the reduction in the
recovered luminosity of Ly$\alpha$ emitting
galaxies is due to aperture photometry losses.
A significant fraction of Ly$\alpha$ photons
are scattered by the circumgalactic
and Intergalactic medium,
forming a diffuse low surface brightness halo. This radiative transfer
effect was discussed 
in detail in Cantalupo et al. 2005 (see also, e.g.,
Zheng et al. 2011, Barnes et al. 2011).
On average, SExtractor aperture photometry
only recovers about 20\% of the 
total luminosity. This is in agreement
with the recent observational estimate
of Steidel et al. (2011) from a stacking analysis.
Once the quasar radiation
is turned on, in the simulation, the Ly$\alpha$ luminosities change, as shown by the solid red dots.
Clearly, fluorescent boosting 
makes the simulated galaxies ``overluminous'' with
respect to their SFR, increasing the
number of detectable objects.
This is due both to the increased emissivity
from recombination radiation and,
to a lesser extent, to 
the reduced Ly$\alpha$ scattering in the
more ionized circum-galactic medium.
In particular, the simulation suggests that
the steepening of the observed luminosity function
is due to galaxies that would
otherwise be one order of magnitude fainter in Ly$\alpha$.
Our survey strategy gives thus the possibility to select 
and study intrinsically faint galaxies at high-redshift
at an incomparably deep level.  This is similar in some respects to the use of gravitational
lenses, although in our
case only the Ly$\alpha$ emission is ``boosted''.
In particular, we should have reached similar levels
to the ultra-deep blind Ly$\alpha$ spectroscopic search
of Rauch et al. 2008, although a direct comparison
is complicated by the very different observational
techniques.

An interesting point is that in the ``QSO-ON'' simulation, we see sources (the
red dots with arrows) that do not have any associated star formation at the simulation
resolution limit for star-formation of about $0.01$ M$_{\odot}$ yr$^{-1}$.
These objects are all below a luminosity of about
10$^{42}$ erg s$^{-1}$ and represent about 20\% of the total
number of sources above our observational
threshold. Their luminosities and relative abundances
are remarkably similar to the observed
objects that did not have continuum-detection. 
A closer look at these simulated sources reveals
that they are compact dense clouds, proto-galaxies 
in the transition phase between the IGM and ISM, 
with peak densities
that are not yet large enough to currently produce and sustain
significant star formation
\begin{footnote}{
From a numerical point of view, this means that
their peak density was always below the assumed
star formation threshold in the simulations of 
$n_{\mathrm{ISM}}=1$ atoms cm$^{-3}$.
}\end{footnote}.

Typically located within dense knots in IGM filaments,
these clouds will eventually increase their gas density
because of cosmological gas infall
and start to form stars. 
Such a ``proto-galactic phase'' for gas rich
and dense clouds might recently have been detected
from the study of high column density absorption systems such 
as LLSs (Fumagalli, O'Meara \& Prochaska 2011) and
among metal-poor DLAs (Penprase et al. 2010; Cooke et al. 2011;
Cooke, Pettini \& Murphy 2012; but see Carswell et al. 2012).
Unfortunately, the lack of spatial and morphological information
intrinsic to these one-dimensional absorption studies does not allow us
to determine whether these clouds
are truly isolated systems or part of larger galactic gas reservoirs, 
either circum-galactic (e.g., streams-like features) or
within galaxies.
Given the statistical evidences for fluorescence in our sample and the
support from simulations, our data provides 
- \emph{in emission} - 
strong indications for
the existence 
of such compact and isolated 
proto-galactic or ``dark'' clouds,
opening up a new observational window for the study of the
early phases of galaxy formation.

How are these emitters then related to QSO absorption line systems? 
In Figure \ref{NHILum} we show the relation between
the Ly$\alpha$ luminosity of our simulated 
sources and the \emph{maximum} HI column density
\emph{before} the quasar is turned on. 
The latter is defined as the maximum value
among the line-of-sight that intersects
any pixel with detectable Ly$\alpha$ emission.
We do not simulate the formation of
molecular hydrogen, therefore
the reported HI values are likely to be an overestimate,
especially at N$_{\mathrm{HI}}>10^{21}$ cm$^{-2}$.
As expected, Ly$\alpha$ emitting galaxies
in the QSO-OFF simulation
are only associated with the highest N$_{\mathrm{HI}}$ systems,
i.e, N$_{\mathrm{HI}}>5\times10^{20}$ cm$^{-2}$,
above the DLA threshold ($2\times10^{20}$ cm$^{-2}$).
This is in a sense by construction, since we only allowed
star-formation to take place above a given density threshold
(1 atom cm$^{-3}$)
following the Schmidt-Kennicutt law (see Appendix B).
Indeed, this density threshold translates 
into a critical
column density (given by the Toomre instability criterion )
that in local galaxies is observed to range
between $5\times 10^{20}$ cm$^{-2}$ and $2\times
10^{21}$ cm$^{-2}$ (Kennicutt 1998).

Once the quasar is turned on, we notice that
the overall boost in luminosity still seems
to correlate with the initial HI column
density of the quasar-off simulation.
This is because higher column density
systems are typically larger, more massive 
and denser. The proto-galactic clouds
without star-formation 
start to appear at values of N$_{\mathrm{HI}}$ around or below
the simulation critical SF threshold ($5\times10^{20}$ cm$^{-2}$). 
In particular, the simulation suggests that
the objects fluorescently illuminated by the quasar
have typical N$_{\mathrm{HI}}$ - when the quasar is off - one order of
magnitude lower than the sources detectable by similarly deep ``blank-field'' surveys.
Given our current sensitivity limits,
we are able to detect only the high-end tail 
of the N$_{\mathrm{HI}}$ distribution associated with
these clouds.  Deeper surveys, e.g. with integral field spectrographs, could reveal many more such systems.

\begin{figure}
\begin{center}
\includegraphics[totalheight=0.3\textheight]{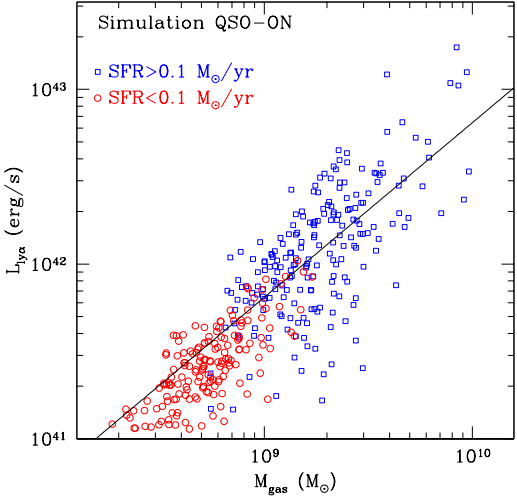}
\caption{The relation between the cloud gas mass and Ly$\alpha$ luminosity
within our simulation with the quasar turned on. 
We have included only gas with $T<5\times10^4$K.
The solid line is the analytical expected relation (eq.\ref{Mgas}) for fully ionized objects without star formation,
derived from the halo cosmological scaling laws and assuming a fixed
gas temperature (see text for details).
This relation gives a good approximation for the expected
mass-luminosity function of fluorescently illuminated gas clouds with
absent or little star-formation, as derived from our radiative transfer simulation. 
Both analytical arguments and the numerical simulation suggest therefore that 
the gas mass of the detected protogalactic candidates would, if it is mostly ionized,
be of the order of 10$^9$ M$_{\odot}$ or larger.  
}
\label{MgasLum}
\end{center}
\end{figure}

\subsection{The gas mass and star formation efficiency of protogalactic haloes}\label{gasmasssec}

If we can assume that there is negligible Ly$\alpha$ emission produced by internal
star formation, then we can directly convert the Ly$\alpha$
luminosity produced by HII recombination into an estimate of the 
total HII mass.
Indeed, in the absence of dust absorption and HI collisional excitations
(in other words, if the gas is mostly ionized), then
the total Ly$\alpha$ luminosity is simply given by (e.g.,
Cantalupo et al. 2008):
\begin{equation} \label{L1}
L_{\mathrm{ly\alpha}}=\int_{V} n_{\mathrm{e}}\cdot n_{\mathrm{HII}}\cdot 
\alpha^{\mathrm{eff}}_{\mathrm{ly\alpha}}(T) h\nu_{\mathrm{ly\alpha}} \mathrm{d}V ,
\end{equation}
where $n_{\mathrm{e}}$, $n_{\mathrm{HII}}$,
$\alpha^{\mathrm{eff}}_{\mathrm{ly\alpha}}(T)$ and
$h\nu_{\mathrm{ly\alpha}}$
 are, respectively, the
electron number density, the proton number density, the
effective Ly$\alpha$ emission coefficient (see
Cantalupo et al. 2008), and the Ly$\alpha$ photon energy. Assuming a uniform temperature 
and ionization fraction within the
cloud, we can rewrite eq.\ref{L1} as:
\begin{equation} \label{L2}
L_{\mathrm{ly\alpha}}=\chi_{\mathrm{e}}
\alpha^{\mathrm{eff}}_{\mathrm{ly\alpha}}(T) h\nu_{\mathrm{ly\alpha}}
\int_{V} n_{\mathrm{HII}}^{2}\mathrm{d}V ,
\end{equation}
where, $\chi_{\mathrm{e}}$ is the electron fraction ($\sim1.16$ if H
and He are fully ionized). Using the definition of clumping factor:
\begin{equation}\label{ClumpingF}
C=\frac{<n_{\mathrm{HII}}^2>}{<n_{\mathrm{HII}}>^2}=
\frac{\int_{V} n_{\mathrm{HII}}^2\mathrm{d}V}{\left[\int_{V}
    n_{\mathrm{HII}}\mathrm{d}V\right]^2}\times V,
\end{equation}
we thus obtain:
\begin{equation}\label{L3}
L_{\mathrm{ly\alpha}}=\chi_{\mathrm{e}}
\alpha^{\mathrm{eff}}_{\mathrm{ly\alpha}}(T)
h\nu_{\mathrm{ly\alpha}}
\cdot C\cdot
  V^{-1} M_{\mathrm{HII}}^2 . 
\end{equation}
For a spherical, fully ionized cloud with radius $R$, gas mass
$M_{\mathrm{gas}}$ and fixed temperature in the range $10^4 K<T<10^{4.7}K$, 
eq.\ref{L3} becomes:
\begin{equation}\label{L4}
L_{\mathrm{ly\alpha}}\sim
1.4\times10^{43} \mathrm{erg/s} \cdot C \cdot
\left[\frac{2\cdot 10^4\mathrm{K}}{T}\right]
\left[\frac{R}{\mathrm{pkpc}}\right]^{-3}
\left[\frac{M_{\mathrm{gas}}}{10^9M_{\odot}}\right]^2 ,
\end{equation}
where pkpc denotes physical kpc.
With the knowledge of the cloud size we could therefore derive its
gas mass from the Ly$\alpha$ luminosity. 

Unfortunately, our proto-galactic
candidates are essentially unresolved in our ground-based
image. 
In absence of this information, 
we can assume that the gas is settled into a disk at the center of its
dark matter halo and derive a proxy for the object size from the 
cosmological scaling relations between virial radius and mass 
(e.g., Mo, Mao \& White 1998). In particular,
assuming that the disk size and gas mass are fixed fractions
(respectively, $f_r$ and $f_g$) of the dark-matter halo virial radius
and total mass, we have (at $z=2.4$):
\begin{equation}\label{R1}
R\sim 2.7 \mathrm{pkpc} \left[\frac{M_{\mathrm{gas}}}{10^9
    M_{\odot}}\right]^{1/3} \left[\frac{f_g}{0.05}\right]^{-1/3}\left[\frac{f_r}{0.1}\right],
\end{equation}
where we have used the same values of the disk to total mass ratio
($f_g$) and disk to virial radius ratio ($f_r$) as in Mo et al. (1998).
It should be noted for consistency that eq.\ref{R1} implies that 
a disk with $M_{\mathrm{gas}}\sim10^{9}$ $M_{\odot}$
would have a size of less than 0.7'' at this redshift and therefore it
would appear unresolved in our image.
Using eq.\ref{R1} as a proxy for the cloud size (and dropping the dependency on $f_g$ and $f_r$ for
simplicity) 
we can thus rewrite eq.\ref{L4} as:
\begin{equation}\label{Mgas}
M_{\mathrm{gas}}\sim 1.4\times 10^9 M_{\odot}
  \left[\frac{L_{\mathrm{ly\alpha}}}{10^{42} \mathrm{erg}\
        \mathrm{s}^{-1}}\right]
\left[\frac{T}{2\times10^{4}K}\right]\cdot C^{-1} .
\end{equation}
We stress that the mass derived with this approach 
is representative of the cloud gas mass only for systems that are highly
ionized by the quasar radiation and with a given temperature. In the other cases, eq.\ref{Mgas} 
underestimates the value of $M_{\mathrm{gas}}$, 
unless there is a significant Ly$\alpha$ emission contribution from HI
collisional excitation, e.g. from cooling. However, highly ionized, optically thick objects
may be distinguished from cooling clouds using
the Ly$\alpha$ spectral and SB profile (see, e.g. Cantalupo et
al. 2005, Dijkstra et al. 2006, Cantalupo et al. 2007)
or with the help of other diagnostic lines such as HeII 1640$\mathrm{\AA}$
(e.g., Yang et al. 2006, Scarlata et al. 2009). 
  Future spectroscopic follow-up and space-based imaging of our sources may therefore help to understand if
eq.\ref{Mgas} is applicabile in these objects. In the current absence of this information, however, 
we can use our radiative transfer simulation to have an idea of the expected
mass-luminosity relation for a variety of objects with different
ionization fractions and temperature distributions.    

 In Fig. \ref{MgasLum} we show the relation between the cloud gas mass
 and its Ly$\alpha$ luminosity in our radiative transfer simulations, once the quasar
 is turned on. We have considered only gas with $T < 5\times10^{4}$
 K - the main component producing Ly$\alpha$ photons from
 recombination - and lying within the aperture radius given by
 SExtractor photometry. 
We divide the sample according to the SFR of the object. For clouds with
 SFR$ < 0.1 M_{\odot}$ yr$^{-1}$ (red circles), the Ly$\alpha$ luminosity is mostly
 powered by quasar radiation. These systems appear to follow well the analytical
 trend with mass. As expected, eq.\ref{Mgas} tends to underestimate
 the total gas mass, although we note that a higher temperature than
 $2\times10^{4}$ K 
would help to recover the value obtained in the simulations.

From eq.\ref{Mgas} and with the help of the radiative transfer simulation, we can
therefore derive an estimate of the gas mass of those Ly$\alpha$ sources
without continuum detection and with EW$_0>240\mathrm{\AA}$. Their
typical Ly$\alpha$ luminosity are in the range $0.5-1\times10^{42}$
erg s$^{-1}$ implying therefore a gas mass of around
$M_{\mathrm{gas}}\sim10^9 M_{\odot}$, or higher if they are not mostly
ionized by the quasar radiation. We can derive a constraint on the
average SFR of these systems from the non-detection in the V-band
stack (see Fig.\ref{Stack}) at the level of
$V(1\sigma)\sim30.3$. We convert this observed magnitude limit
(corresponding to rest-frame wavelength of about 1600$\mathrm{\AA}$)
 into a SFR upper limit using Starburst99 (Leitherer et al. 1999) and
 Oti-Floranes \& Mas-Hesse (2010). Assuming an extended burst of
 duration 250Myr, Salpeter IMF (1-100 $M_{\odot}$) and $E(B-V)=0$ we
 obtain a SFR$<10^{-2} M_{\odot}$ yr$^{-1}$. This implies
 a Star Formation Efficiency (SFE) for these objects of 
SFE$\equiv\mathrm{SFR}/M_{gas}<10^{-11}$ yr$^{-1}$.

This very low SFE is several times lower than the typical SFE of dwarf
galaxies in the local Universe (e.g., Geha et al. 2006).
It is also 200 times lower than the SFE of typical main sequence galaxies at $z \sim 2$ which has been estimated to be around 
SFE$ \sim 2 \times10^{-9}$ yr$^{-1}$
(Daddi et al. 2010, Genzel et al. 2010).
Although we currently have only upper limits on the cloud sizes,
it is interesting to estimate where our sources would lay with respect
to the Kennicutt-Schmidt (KS) relation (Kennicutt 1998) between SFR per unit area
($\Sigma_{\mathrm{SFR}}$) and gas surface density ($\Sigma_{\mathrm{gas}}$). From the
estimated gas mass $M_{\mathrm{gas}}\sim10^9 M_{\odot}$, the SFR upper limit and assuming
that our objects follow the cosmological scaling relation between mass
and radius (eq. \ref{R1}), we would obtain an average 
$\Sigma_ {\mathrm{gas}}\sim44\ M_{\odot}$ pc$^{-2}$
and $\Sigma_{\mathrm{SFR}}<4\times10^{-4}\ M_{\odot}$ yr$^{-1}$
  Kpc$^{-2}$. This value of $\Sigma_{\mathrm{SFR}}$ is about 100 times
lower than the expected value from the local KS relation but it is
consistent with the metallicity-dependent models of Krumholz, McKee
and Tumlinson (2009) and Gnedin \& Kravtsov (2011) that predict a substantial
steepening of the KS relation at $\Sigma_ {\mathrm{gas}}\sim50-100 M_{\odot}$
for very low metallicity gas ($Z<0.01Z_{\odot}$) at high redshift
(see also Rafelski, Wolfe \& Chen 2011).

\begin{figure}
\begin{center}
\includegraphics[totalheight=0.25\textheight]{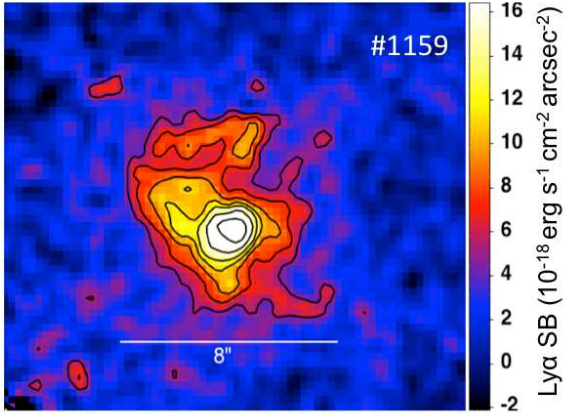}
\caption{Continuum subtracted NB image of \#1159 revealing
the extended, filamentary emission surrounding the central
source. Contours show isophotal surfaces (1$\sigma$ intervals,
starting from 2$\sigma$). Size and luminosity of this emission are
compatible with the circumgalactic medium illuminated by the quasar
(see text for details). Note that 8'' corresponds to about 66 physical
kpc at z=2.4, i.e. the virial diameter of a halo with total mass
$M\sim4\times10^{11} M_{\odot}$. 
}
\label{cs1159}
\end{center}
\end{figure}

\subsection{Cold Circum-Galactic Medium}\label{CGMsec}

A different category of objects is represented by the more luminous, continuum-detected
sources with EW$>240\mathrm{\AA}$, which show evidence of extended
emission (left-hand panels in Figure \ref{LAEFig}). 
These are among the brightest Ly$\alpha$ sources in our sample
and therefore it is not surprising that we detected
some stellar continuum associated with them. 
Their extended emission has a surface brightness at a level
of $0.5-1\times 10^{-17}$ erg s$^{-1}$ cm$^{-2}$  arcsec$^{-2}$,
a value compatible with the quasar induced fluorescence 
suggested by the high EW. 
Their luminosities and low number density suggests, on the basis of our simulations,
an association  
with the most massive systems in our surveyed volume,
i.e. with objects
with halo masses between 10$^{11}$ and 10$^{12}$
$M_{\odot}$.  Several theoretical and numerical
studies have suggested that these haloes 
are able to retain 
dense and cold gas reservoirs in the form
of filaments in the circum-galactic medium (e.g., Dekel \& Birnboim
2006, Ocvirk \& Teyssier 2008, Dekel et al. 2009,
Keres et al. 2009, Fumagalli et al. 2011, van de Voort \& Schaye 2012). 
At the densities and temperatures predicted by these
models, these circum-galactic filaments could be
detectable in fluorescent emission in the proximity
of a bright quasar. 
Note that this is a different mechanism
with respect to Ly$\alpha$ cooling emission - 
suggested as a possible origin
for extended Ly$\alpha$ structures like
Ly$\alpha$ blobs, \emph{away} from bright quasars
(e.g., Haiman, Spaans \& Quataert 2000, Fardal et al. 2001, Dijkstra \& Loeb 2009, Goerdt et al. 2010;
but see Furlanetto et al. 2005, Faucher-Giguere et al. 2010; Rosdahl
\& Blaizot 2011). Mostly due to HII recombinations, quasar fluorescent emission
is much less sensitive to the gas temperature state (see e.g., Fig. 1
in Cantalupo et al. 2008) and therefore it is much easier to estimate
the expected Ly$\alpha$ surface brightness (see Appendix B for further discussion).
It is intriguing that
one of the brightest objects in our sample, \#1159, indeed shows some evidence
for possibly filamentary emission features (Fig. \ref{cs1159})
extending over 8'' in the sky. At $z=2.4$, this corresponds to a projected size of 
66 physical kpc, i.e. the virial diameter of a halo with total mass $M\sim4\times10^{11}
M_{\odot}$. Without kinematical information, it is however difficult
to determine if these extended features are associated
with infall or outflows. 

Other mechanisms can produce filamentary structures. In fact, in our simulations,
the most common sign of filamentary CGM emission
is associated with tidal features,
e.g., dense gas stripped out of satellite galaxies,
rather than the cold infall of gas.  This could be the origin of the extended emission observed in our images. In particular
the observed ``bridge'' between two of our bright extended 
sources, \#2411 and \#2433 might fall into this category (see also
Rauch et al. 2011).
Further theoretical and observational studies
focusing on these sources, e.g.,
with the help of Integral Field Spectroscopy,
might help reveal the detailed properties 
of the detected extended emission.   

While much remains to be learnt about these systems, the key point is that our narrow-band survey has demonstrated
the potential of finding and mapping
the extended, cold circum-galactic
medium around galaxies thanks to
fluorescent boosting of the emission. 
This mechanism opens up
a new window for the study of
this important gas reservoir
of high-redshift galaxies
and to unraveling how galaxies
receive their gas. 

\section{Uncertainties and Limitations}\label{uncertaintiessec}

\subsection{Foreground and AGN contamination}\label{fgSec}

What fraction of our Ly$\alpha$ candidates
could be foreground line emitting galaxies
and how do this affects our results?
Typically, the primary sources of contamination
for high-redshift NB Ly$\alpha$ surveys
are foreground galaxies with
strong [OII] 3726-3729$\angs \ $
emission line doublet.
For Ly$\alpha$ surveys at $2<z<3$, corresponding to $0.1<z<0.3$ 
for [OII], the typical fraction
of foreground contamination
is typically less than 5\% (e.g., Venemans et al. 2007),
because of the much smaller volume and smaller EW
associated with [OII] emitters.
From our data, we cannot directly estimate
the foreground contamination
in our sample because of the
lack of spectroscopic confirmation
for the whole sample. The spectroscopic
follow-up of about twenty among the
brightest objects in our sample 
(Cantalupo et al., in prep.)
clearly shows that they are Ly$\alpha$
emitters but the intrinsic faintness
for the rest of our sample makes
their direct confirmation more difficult.

Although foreground contamination could increase the number of detectable
objects (affecting the luminosity function), 
it cannot produce our skewed EW distribution
towards higher values. On the contrary, 
given their smaller EW,
foreground objects would be expected to
produce the opposite effect. Moreover, 
they would not show any correlation
between their luminosity and the
position of the quasar. Therefore, 
we believe that the presence of
foregrounds in our sample 
does not affect our conclusions.

Unlike star-powered galaxies, active galactic nuclei (AGN) are
able to produce EW well above the
canonical $240\angs \ $ value and 
might thus be ``confused'' with
fluorescent emission. AGN can be
identified by spectroscopic features
(e.g., broad emission lines and the presence
of the CIV doublet) or from X-ray 
coincident detections. Previous surveys have
shown that AGN contamination is
always very low, of the order of few
percent (e.g., Guaita et al. 2010). 
It is in principle conceivable that there might be an increased number of AGN
around our ultra-luminous quasar.  Although clustering effects
would increase the number of galaxies around the quasar, to our knowledge 
there are no indications in the literature for the increase, 
by a factor of at least five, in the AGN fraction in such environments,
which would be necessary to explain our results in terms of AGN.
This is corroborated by our preliminary
spectroscopic confirmation for the
bright part of our sample (Cantalupo et al., in prep.):
only one object out of about twenty 
shows the presence of CIV emission.
Therefore, although AGN might
contribute to increasing the number
of high EW objects in our sample, it is
very unlikely that they could account
for observed difference with the ``blank-fields''.

\subsection{Collisionally excited Ly$\alpha$}

Ly$\alpha$ produced by collisional excitation via energetic electrons,
commonly called ``Ly$\alpha$
cooling'', is a mechanism that is in principle
able to produce strong Ly$\alpha$ emission
without associated star formation, i.e.
similarly to quasar fluorescence. 
Unfortunately, collisionally excited
Ly$\alpha$ emission has an exponential dependence
on gas temperature (see e.g., Fig.1 in Cantalupo
et al. 2008), which makes accurate predictions
within current hydro-simulations 
very challenging (see Appendix B for discussion).

For the primary purpose of this study - finding proto-galactic
clouds and direct imaging of the cold CGM - 
it does not really matter if Ly$\alpha$
photons are produced by quasar fluorescence
or by cooling due to gravitational collapse.
From an observational
point of view, Ly$\alpha$ cooling should be equally detectable away from bright quasars.
If our sources are in fact powered by cooling
instead of fluorescence,
then the ``blank-field'' surveys at $2<z<3$ should
already have detected such objects, and we should not see any particular
difference between, e.g. our EW distribution
and the one obtained by these surveys.
On the other hand, cooling Ly$\alpha$ emission
could explain why ``blank-field'' surveys
\emph{do} detect some objects with high EW away from
bright quasars, including the large Ly$\alpha$
blobs. These sources would also then be present
within our survey. However, given their low space
density (e.g., Matsuda et al. 2004) 
we would expect that they are responsible for only
a small fraction of our sample of high EW sources.

\subsection{Quasar effects}

The ionizing radiation from the quasar
has the positive effect, for us, of 
increasing the ionization state, and thus
the recombinant Ly$\alpha$ emission, of the gas. 
However, the same radiation could also have
an important effect on the galaxy-formation
process itself around the quasar,
e.g., reducing the gas cooling rate (see Cantalupo 2010)
and thus effectively limiting gas
accretion onto (proto-)galaxies and suppressing their SFRs.
In principle, from the continuum properties of the sources in our survey
it would be possible to search for a general star-forming
deficit relative to the ``blank-fields''. However,
the intrinsic faintness of our objects
plus uncertain clustering effects
would make this task
very difficult in practice.

Nevertheless, 
let us assume for the moment that the galaxy-formation
process would be ``slowed-down''
for a gas cloud 
while it is
``illuminated'' by the quasar, e.g. if the gas
is completely ionized. 
In this case, the overall effect on the galaxies would depend 
on the quasar emission history compared to the galaxy-formation
timescales. Within the current paradigm
of quasar black-hole accretion
the quasar luminous phase is expected to have a relatively
short duration (e.g. a few tens of Myrs) followed
by a much longer phase (e.g., a few Gyrs) of lower ultraviolet output 
(e.g., Di Matteo et al. 2008; see also Martini 2004 for a review).
The dense gas, which should be associated with
quasar fluorescent emission has a recombination
time scale is very much shorter, of the order
of $10^5-10^6$ yrs, than the quasar lifetime, and of cosmological timescales.  This means that any
{\it previous} bright phase of the quasar will be quickly ``forgotten'' once the quasar turns off,
and proto-galactic
clouds and galaxies should be able to recover
their ``normal'' evolution before the current
phase. 
At the same time, the star formation timescale probed by our deep V band image (centered on
$\lambda\sim\mathrm{1600\AA}$) is at least 
200 Myrs (e.g., Leitherer et al. 1999), one order of magnitude larger than
the expected quasar luminous phase. Therefore it is unlikely that the {\it current} 
bright phase is the cause of the apparent lack of star formation
in our proto-galactic clouds candidates. 
Of course, we cannot exclude
that our quasar had a very peculiar emission
history. Future observational campaigns
will be crucial to increase the
the sample to a larger
number of quasars and therefore
reduce these uncertainties.
  
\section{Summary and Conclusions}\label{summarysec}

We have presented a deep survey for fluorescent
Ly$\alpha$ emission in a large cosmological volume
($\sim 5\times 10^3$ comoving Mpc)
centered on the hyperluminous $z = 2.4$ quasar HE0109-3518.
With the help of a custom-made, 
narrow-band filter mounted on the VLT/FORS spectrograph,
we were able to photometrically select
98 Ly$\alpha$ candidates down to a
detection limit of about $4\times 10^{-18}$ erg s$^{-1}$
cm$^{-2}$ (corresponding to $L_{Ly\alpha}\sim 2\times 10^{41}$
erg $s^{-1}$ at $z=2.4$).

The properties of these Ly$\alpha$ sources in terms of their:
i) equivalent width distribution,
ii) shape of the luminosity function 
and,
iii) the variation of their luminosity with projected
distance from the quasar, 
all indicate that our
sample is consistent with having much of the Ly$\alpha$ emission originating in
fluorescent reprocessing of quasar radiation. 

In particular, the main evidence supporting
the fluorescent origin of our sample are:
\begin{itemize}
\item
The large number of objects with rest-frame 
EW$_0>240\angs \ $ (18 sources out of 98), 
commonly assumed 
to be the limit for Ly$\alpha$
emission powered by Population II stars 
(e.g., Schaerer 2002).
Independent of luminosity, the fraction of such objects appears to be about 
20\%,
which is about a factor of four larger
than ``blank-field'' surveys (e.g.,
Ciardullo et al. 2012, Grove et al.
2009) or surveys around radio-galaxies
(Venemans et al. 2007) at similar redshifts. 
Moreover, the whole shape
of our equivalent width distribution is skewed toward 
higher values, with an exponential scale
that is about double that found in the ``blank-fields''.
\item The  steeper faint-end of our LF with respect
to the ``blank-fields'' surveys with a corresponding increase in the source
number density by a factor of about five compared to these studies. 
The steepening of the LF is well 
reproduced by a radiative transfer
simulation of fluorescent emission around quasars.
The simulation suggests that  the difference with the ``blank-field''
surveys may be fully explained by fluorescent boosting of gas-rich
but intrinsically faint Ly$\alpha$ emitters. 
\item The increase in the average luminosity
of the sources at smaller (projected)
distances from the quasar.
Although this effect could also
be caused by clustering of bright
sources around the quasar,
our simulations suggest that the observed
trend is very similar to the theoretical
expectations for fluorescent emission.
\end{itemize}

Fluorescent emission gives us a unique opportunity to
directly image, in emission, dense, high-redshift gas
independent of any associated star formation.
In particular, we identify three categories of objects associated
with fluorescent emission in our study:
\begin{itemize}
\item \emph{Proto-galactic or ``dark'' clouds, i.e. gas-rich
objects with little or no associated
star-formation}.  Candidates for these sources are represented by the
12 objects without detectable 
continuum and with lower limits on their EW
higher above $240\angs \ $. They are
typically compact or unresolved
with $L\lesssim 10^{42}$ erg s$^{-1}$.
A V-band image stack of these objects did not show any continuum detection above
1$\sigma$ ($V\sim30.3$), implying a combined constraint on their EW of 
EW$_0>800\mathrm{\AA}$ (1$\sigma$). This effectively rules out Ly$\alpha$
powered by internal star formation.
A comparison with our radiative transfer
simulations suggests that these sources
would have maximal HI column
densities around or slightly below
the DLA threshold ($2\times 10^{20}$ cm$^{-2}$) if they were
observed \emph{away} from the quasar.
These may therefore be the analogues of absorption systems
which have also been suggested to probe 
such a ``proto-galactic phase''
(e.g., Fumagalli et al. 2011, Cooke et al. 2011).
Observing these systems in emission - rather than in absorption - 
allows us to readily discriminate between truly isolated clouds and  
sub-components of larger, galactic or circum-galactic 
gas reservoirs. 
Our narrow-band imaging 
provides indications for the existence of compact and isolated
proto-galactic or ``dark'' clouds
opening up a new observational window for the study of the
early phases of galaxy formation.
Assuming that these objects are almost fully ionized by
the intense quasar radiation, we can directly derive an estimate or
their total gas mass, namely $M_{\mathrm{gas}}\sim10^9 M_{\odot}$.
From the non-detection in the V-band stack, we can furthermore estimate an upper
limit on their average star formation rate of SFR$<10^{-2}
M_{\odot}$ yr$^{-1}$. This implies that their star formation
efficiency is lower than 
SFE$<10^{-11}$ yr$^{-1}$ (or,
equivalently, their gas consumption time is larger than 100 Gyr),
several times below the SFE of gas-rich dwarf galaxies in the
local Universe (e.g., Geha et al. 2006) and two hundred times lower
than in typical massive star-forming galaxies 
at these redshifts (Daddi et al, 2010, Genzel et al 2010).
These objects are therefore
the best candidates for proto-galactic clouds or ``dark'' galaxies at
high-redshift suggested by recent theoretical studies 
(e.g., Bouch\'e et al. 2010, Gnedin \& Kravtsov 2010, Krumholz \&
Dekel 2011, Kuhlen et al. 2012).

\item \emph{Intrinsically faint Ly$\alpha$
emitting galaxies fluorescently
boosted above our detection threshold}.
As suggested by our 
radiative transfer simulations, 
these sources are the main component
producing the observed
e-folding scale increase in the
EW distribution and the steepening
of the LF. In particular, their implied 
SFRs are about one order of magnitude
smaller than the values we could
have detected in absence of quasar
fluorescence.
\item \emph{Extended, cold circum-galactic medium emission}.
At least three of the six bright sources
with EW$_0>240\angs \ $and continuum detection
in our sample show extended, possibly filamentary
fluorescent emission. In particular, one object 
(\#1159)
presents signs of several filaments connected
to a bright central source.
The size and morphology of these
structures are compatible with 
the expectations from theoretical
and numerical models 
for cold streams feeding high-redshift
galaxies (e.g., Dekel et al. 2009). Our current data
and the lack of kinematical information
cannot however rule out a different
origin for this cold and dense
circum-galactic gas, e.g., 
tidal stripping from gas-rich
satellite galaxies (as commonly
seen in our simulations).
\end{itemize}

After several years of
attempts to detect 
fluorescent Ly$\alpha$ emission,
our results demonstrate the
potential of a carefully 
designed (very-)narrow-band 
survey
around bright quasars
to discover and study this new category
of sources. 
Several important factors have contributed  to this success.
As we have shown,  a large survey volume, comparably deep  narrow-band and 
continuum imaging, as well as  sophisticated numerical models including radiative transfer effects,
are fundamental to detect, recognize and interpret fluorescent emission.
Still, the success of such a fluorescence survey
might depend on the emitting history and
properties of the source that provides
the ``illumination'', i.e. the quasar.
In this respect, our survey suggests that 
for \QSOtwo\ at least, the proximity
effect extends on much larger angular
scale than just our line-of-sight,
the intrinsic limit of absorption-line studies. 
Object by object variations, including
clustering and cosmic variance, might however
play an important role and observational campaigns targeting
a larger number of quasars will thus be important in the future. 

By stimulating fluorescent emission in neutral gas, the quasars are literally 
shedding light on the hitherto
obscure early stages of
galaxy-formation and on the
way in which galaxies acquire
their gas. 
With the present study,
we make a crucial step
towards revealing and 
understanding how
to take full advantage
of this unique opportunity.

\section*{Acknowledgments}

We are grateful to the staff of Paranal, Chile for their excellent
support. 
We thank the referee for a very detailed and careful review of the manuscript
and for constructive questions and comments.
 SC thanks
J. Xavier Prochaska for carefully reading an earlier version of the manuscript and providing
suggestions that improved the presentation of this work and George
Becker, Mattew Hayes, 
Piero Madau, Richard McMahon,
Michael Rauch, Debora Sijacki, Dan Stark and Gabor Worsek 
for many very useful discussions. 
The hydrodynamical simulations used in this work were
performed using the Darwin Supercomputer of the University
of Cambridge High Performance Computing Service
(http://www.hpc.cam.ac.uk/), provided by Dell Inc. using Strategic
Research Infrastructure Funding from the Higher Education
Funding Council for England.
SC acknowledges support from NSF grants AST-1010004.
This  research was  supported in part by the National Science Foundation under 
Grant No. PHY11-25915.

\appendix

\onecolumn

\section{List of continuum detected objects with EW$_0< 240 \mathrm{\AA}$ }
\begin{center}
\tablefirsthead{%
\multicolumn{9}{c}{Table A1: Position and properties Ly$\alpha$ candidates with continuum detection and EW$_0< 240 \mathrm{\AA}$ (see Table 2)} \\
\hline
\hline
\multicolumn{1}{c}{Id} &
\multicolumn{1}{c}{RA} &
\multicolumn{1}{c}{DEC}&
\multicolumn{1}{c}{NB} &
\multicolumn{1}{c}{V}  &
\multicolumn{1}{c}{B}  &
\multicolumn{1}{c}{F} &
\multicolumn{1}{c}{L$_{Ly\alpha}$} &
\multicolumn{1}{c}{EW$_{0}$}
 \\
\multicolumn{1}{c}{ } &
\multicolumn{1}{c}{(J2000)} &
\multicolumn{1}{c}{(J2000)}&
\multicolumn{1}{c}{(AB)} &
\multicolumn{1}{c}{(AB)}  &
\multicolumn{1}{c}{(AB)}  &
\multicolumn{1}{c}{10$^{-17}$ erg s$^{-1}$ cm$^{-2}$} &
\multicolumn{1}{c}{10$^{42}$ erg s$^{-1}$} &
\multicolumn{1}{c}{$\mathrm{\AA}$}
\\
\hline}
\tablehead{%
\multicolumn{9}{c}{Table A1 (\emph{continued})} \\
\hline
\hline
\multicolumn{1}{c}{Id} &
\multicolumn{1}{c}{RA} &
\multicolumn{1}{c}{DEC}&
\multicolumn{1}{c}{NB} &
\multicolumn{1}{c}{V}  &
\multicolumn{1}{c}{B}  &
\multicolumn{1}{c}{F$_{Ly\alpha}$} &
\multicolumn{1}{c}{L$_{Ly\alpha}$} &
\multicolumn{1}{c}{EW$_{0}$}
 \\
\multicolumn{1}{c}{ } &
\multicolumn{1}{c}{(J2000)} &
\multicolumn{1}{c}{(J2000)}&
\multicolumn{1}{c}{(AB)} &
\multicolumn{1}{c}{(AB)}  &
\multicolumn{1}{c}{(AB)}  &
\multicolumn{1}{c}{10$^{-17}$ erg s$^{-1}$ cm$^{-2}$} &
\multicolumn{1}{c}{10$^{42}$ erg s$^{-1}$} &
\multicolumn{1}{c}{$\mathrm{\AA}$}
\\
\hline}
\tablelasttail{%
\hline
}
\begin{supertabular}{*{9}{c}}

 2433  & 1:11:33.85 & -35:01:12.4 &  23.32$^{+ 0.09 }_{ -0.08 }$ &  25.35$^{+ 0.12 }_{ -0.11 }$ &  26.3 $^{+ 0.11 }_{ -0.1 }$ & 12.22$\pm$0.97 &  5.68$\pm$0.45 &  218$\pm$27  \\
 248  & 1:11:58.01 & -35:06:08.9 &  24.85$^{+ 0.08 }_{ -0.08 }$ &  28.05$^{+ 0.28 }_{ -0.23 }$ & $>$ 28.49  &  2.99$\pm$0.21 &  1.39$\pm$0.10 &  207$\pm$45  \\
 973  & 1:11:40.02 & -35:04:30.7 &  24.77$^{+ 0.08 }_{ -0.08 }$ &  27.88$^{+ 0.30 }_{ -0.24 }$ & $>$ 28.58  &  3.21$\pm$0.23 &  1.49$\pm$0.11 &  192$\pm$45  \\
 558  & 1:11:28.50 & -35:05:22.7 &  25.44$^{+ 0.12 }_{ -0.11 }$ &  28.42$^{+ 0.25 }_{ -0.21 }$ & $>$ 28.89  &  1.73$\pm$0.18 &  0.81$\pm$0.08 &  168$\pm$35  \\
 2232  & 1:11:38.52 & -35:00:39.4 &  25.75$^{+ 0.12 }_{ -0.11 }$ &  28.65$^{+ 0.32 }_{ -0.25 }$ &  28.31$^{+ 0.31 }_{ -0.25 }$ &  1.30$\pm$0.14 &  0.61$\pm$0.06 &  155$\pm$40  \\
 2360  & 1:11:42.06 & -35:00:58.7 &  24.43$^{+ 0.06 }_{ -0.06 }$ &  27.29$^{+ 0.13 }_{ -0.12 }$ &  27.88$^{+ 0.31 }_{ -0.25 }$ &  4.40$\pm$0.24 &  2.04$\pm$0.11 &  149$\pm$15  \\
 1946  & 1:11:38.58 & -35:00:07.4 &  24.19$^{+ 0.10 }_{ -0.10 }$ &  25.38$^{+ 0.13 }_{ -0.12 }$ &  26.68$^{+ 0.14 }_{ -0.13 }$ &  5.48$\pm$0.48 &  2.55$\pm$0.22 &  147$\pm$24  \\
 1198  & 1:11:31.04 & -35:03:55.3 &  23.85$^{+ 0.05 }_{ -0.05 }$ &  26.36$^{+ 0.09 }_{ -0.09 }$ &  26.60$^{+ 0.16 }_{ -0.15 }$ &  7.50$\pm$0.34 &  3.48$\pm$0.16 &  145$\pm$20  \\
 1899  & 1:11:36.67 & -35:02:59.1 &  23.63$^{+ 0.09 }_{ -0.09 }$ &  25.39$^{+ 0.10 }_{ -0.09 }$ &  26.22$^{+ 0.12 }_{ -0.11 }$ &  9.18$\pm$0.73 &  4.27$\pm$0.34 &  144$\pm$18  \\
 2818  & 1:11:51.42 & -35:02:05.8 &  23.38$^{+ 0.08 }_{ -0.07 }$ &  25.25$^{+ 0.07 }_{ -0.06 }$ &  25.96$^{+ 0.09 }_{ -0.09 }$ & 11.56$\pm$0.82 &  5.37$\pm$0.38 &  139$\pm$12  \\
 2965  & 1:11:31.74 & -35:02:20.5 &  25.54$^{+ 0.10 }_{ -0.09 }$ &  28.17$^{+ 0.32 }_{ -0.25 }$ & $>$ 28.46  &  1.58$\pm$0.14 &  0.73$\pm$0.06 &  118$\pm$30  \\
 2958  & 1:11:40.94 & -35:02:22.5 &  24.66$^{+ 0.07 }_{ -0.07 }$ &  26.52$^{+ 0.09 }_{ -0.09 }$ &  27.1$^{+ 0.23 }_{ -0.20 }$ &  3.56$\pm$0.22 &  1.65$\pm$0.10 &  117$\pm$23  \\
 2345  & 1:11:33.62 & -35:01:03.0 &  24.58$^{+0.13 }_{ -0.12 }$ &  25.81$^{+ 0.11 }_{ -0.1 }$ &  26.86$^{+ 0.16 }_{ -0.15 }$ &  3.83$\pm$0.43 &  1.78$\pm$0.20 &  112$\pm$19  \\
 1172  & 1:11:51.25 & -35:03:59.6 &  24.39$^{+0.13 }_{ -0.12 }$ &  25.63$^{+ 0.14 }_{ -0.13 }$ &  26.66$^{+ 0.15 }_{ -0.14 }$ &  4.56$\pm$0.51 &  2.12$\pm$0.24 &  111$\pm$20  \\
 510  & 1:11:35.90 & -35:05:28.9 &  25.89$^{+0.16 }_{ -0.14 }$ &  28.44$^{+ 0.26 }_{ -0.22 }$ &  28.12$^{+ 0.30 }_{ -0.24 }$ &  1.15$\pm$0.16 &  0.53$\pm$0.07 &  110$\pm$25  \\
 1487  & 1:11:36.40 & -35:03:13.5 &  24.70$^{+0.15 }_{ -0.14 }$ &  26.11$^{+ 0.14 }_{ -0.13 }$ &  26.99$^{+ 0.23 }_{ -0.20 }$ &  3.43$\pm$0.44 &  1.59$\pm$0.21 &  109$\pm$25  \\
 2547  & 1:11:37.98 & -35:01:28.6 &  24.29$^{+0.05 }_{ -0.05 }$ &  26.94$^{+ 0.13 }_{ -0.12 }$ &  26.83$^{+ 0.16 }_{ -0.15 }$ &  5.00$\pm$0.23 &  2.32$\pm$0.10 &  107$\pm$17  \\
 2251  & 1:11:29.99 & -35:00:44.7 &  24.95$^{+ 0.15 }_{ -0.14 }$ &  26.38$^{+ 0.14 }_{ -0.13 }$ &  27.2$^{+ 0.22 }_{ -0.19 }$ &  2.72$\pm$0.35 &  1.26$\pm$0.16 &  102$\pm$23  \\
 2804  & 1:11:51.42 & -35:02:03.2 &  24.44$^{+ 0.13 }_{ -0.12 }$ &  26.18$^{+ 0.13 }_{ -0.12 }$ &  26.7$^{+ 0.16 }_{ -0.14 }$ &  4.35$\pm$0.49 &  2.02$\pm$0.23 &  95$\pm$17  \\
 1881  & 1:11:46.41 & -34:59:56.5 &  25.52$^{+ 0.09 }_{ -0.09 }$ &  27.92$^{+ 0.3 }_{ -0.24 }$ &  28.16$^{+ 0.43 }_{ -0.32 }$ &  1.61$\pm$0.13 &  0.75$\pm$0.06 &  94$\pm$22  \\
 2629  & 1:11:41.16 & -35:01:38.1 &  24.06$^{+ 0.10 }_{ -0.09 }$ &  25.73$^{+ 0.09 }_{ -0.08 }$ &  26.28$^{+ 0.11 }_{ -0.10 }$ &  6.18$\pm$0.54 &  2.87$\pm$0.25 &  92$\pm$10  \\
 2594  & 1:11:35.80 & -35:01:29.3 &  24.68$^{+ 0.15 }_{ -0.14 }$ &  27.04$^{+ 0.21 }_{ -0.18 }$ & $>$ 27.99  &  3.49$\pm$0.45 &  1.62$\pm$0.21 &  90$\pm$18  \\
 2496  & 1:11:37.72 & -35:01:17.0 &  25.21$^{+0.10 }_{ -0.10 }$ &  27.27$^{+ 0.17 }_{ -0.15 }$ &  27.33 $^{+ 0.25 }_{ -0.21 }$ &  2.14$\pm$0.19 &  1.00$\pm$0.09 &  73$\pm$18  \\
 567  & 1:11:48.19 & -35:05:22.0 &  25.77$^{+ 0.12 }_{ -0.11 }$ &  27.92$^{+ 0.15 }_{ -0.14 }$ & $>$ 29.14  &  1.28$\pm$0.13 &  0.59$\pm$0.06 &  72$\pm$10  \\
 1837  & 1:11:39.74 & -34:59:52.7 &  24.96$^{+0.08 }_{ -0.08 }$ &  26.92$^{+ 0.11 }_{ -0.10 }$ &  27.05 $^{+ 0.16 }_{ -0.15 }$ &  2.70$\pm$0.19 &  1.25$\pm$0.09 &  71$\pm$11  \\
 2682  & 1:11:42.31 & -35:01:46.1 &  25.18$^{+0.10 }_{ -0.10 }$ &  26.71$^{+ 0.10 }_{ -0.10 }$ &  27.17 $^{+ 0.24 }_{ -0.20 }$ &  2.20$\pm$0.19 &  1.02$\pm$0.09 &  70$\pm$15  \\
 525  & 1:11:58.63 & -35:05:29.8 &  25.80$^{+ 0.12 }_{ -0.12 }$ &  27.89$^{+ 0.24 }_{ -0.2 }$ & $>$ 28.64  &  1.24$\pm$0.13 &  0.58$\pm$0.06 &  68$\pm$14  \\
 1649  & 1:11:41.78 & -35:03:34.9 &  25.90$^{+ 0.14 }_{ -0.13 }$ &  27.99$^{+ 0.35 }_{ -0.27 }$ & $>$ 28.29  &  1.13$\pm$0.14 &  0.53$\pm$0.06 &  68$\pm$19  \\
 1413  & 1:11:45.24 & -35:03:06.6 &  25.26$^{+ 0.08 }_{ -0.08 }$ &  27.35$^{+ 0.18 }_{ -0.16 }$ & $>$ 28.56  &  2.05$\pm$0.15 &  0.95$\pm$0.07 &  67$\pm$10  \\
 1931  & 1:11:28.37 & -35:00:12.4 &  26.51$^{+ 0.19 }_{ -0.17 }$ &  28.59$^{+ 0.45 }_{ -0.33 }$ & $>$ 28.71  &  0.65$\pm$0.10 &  0.30$\pm$0.05 &  67$\pm$24  \\
 2076  & 1:11:32.03 & -35:00:25.9 &  26.81$^{+ 0.17 }_{ -0.16 }$ &  28.88$^{+ 0.36 }_{ -0.28 }$ & $>$ 29.19  &  0.49$\pm$0.07 &  0.23$\pm$0.03 &  66$\pm$20  \\
 1538  & 1:11:46.44 & -35:03:22.4 &  25.14$^{+0.20 }_{ -0.18 }$ &  27.15$^{+ 0.35 }_{ -0.27 }$ & $>$ 28.31  &  2.29$\pm$0.38 &  1.06$\pm$0.18 &  62$\pm$19  \\
 963  & 1:11:29.86 & -35:04:35.7 &  25.24$^{+0.10 }_{ -0.10 }$ &  26.53$^{+ 0.09 }_{ -0.09 }$ &  27.06$^{+ 0.20 }_{ -0.18 }$ &  2.08$\pm$0.18 &  0.97$\pm$0.09 &  60$\pm$11  \\
 791  & 1:11:57.22 & -35:04:53.0 &  26.36$^{+0.16 }_{ -0.14 }$ &  28.34$^{+ 0.23 }_{ -0.19 }$ &  28.18$^{+ 0.29 }_{ -0.24 }$ &  0.74$\pm$0.10 &  0.35$\pm$0.05 &  60$\pm$13  \\
 380  & 1:11:41.15 & -35:05:51.0 &  26.12$^{+ 0.17 }_{ -0.15 }$ &  28.1$^{+ 0.29 }_{ -0.23 }$ & $>$ 28.55  &  0.93$\pm$0.13 &  0.43$\pm$0.06 &  60$\pm$15  \\
 471  & 1:11:53.11 & -35:05:35.9 &  25.72$^{+ 0.12 }_{ -0.11 }$ &  27.69$^{+ 0.18 }_{ -0.16 }$ & $>$ 28.71  &  1.34$\pm$0.14 &  0.62$\pm$0.07 &  59$\pm$10  \\
 1537  & 1:11:46.38 & -35:03:21.6 &  25.21$^{+ 0.19 }_{ -0.17 }$ &  27.15$^{+ 0.35 }_{ -0.27 }$ & $>$ 28.31  &  2.14$\pm$0.34 &  1.00$\pm$0.16 &  57$\pm$17  \\
 683  & 1:11:44.21 & -35:04:59.2 &  23.76$^{ +0.08 }_{ -0.08}$ & 24.71$^{+0.05}_{ -0.05}$ & 25.45$^{+  0.12}_{ -0.11}$ &  8.15$\pm$0.58 &  3.78$\pm$0.27 & 55$\pm$6 \\ 
2004  & 1:11:37.64 & -35:00:15.4 &  25.56$^{+ 0.11 }_{ -0.10 }$ &  27.43$^{+ 0.20 }_{ -0.17 }$ & $>$ 28.38  &  1.55$\pm$0.15 &  0.72$\pm$0.07 &  53$\pm$9  \\
 1574  & 1:11:29.77 & -35:03:24.8 &  25.21$^{+ 0.11 }_{ -0.10 }$ &  27.07$^{+ 0.19 }_{ -0.16 }$ & $>$ 28.05  &  2.14$\pm$0.21 &  1.00$\pm$0.10 &  52$\pm$9  \\
 2714  & 1:11:41.74 & -35:01:49.3 &  26.16$^{+ 0.20 }_{ -0.18 }$ &  27.96$^{+ 0.24 }_{ -0.20 }$ & $>$ 28.59  &  0.89$\pm$0.15 &  0.41$\pm$0.07 &  49$\pm$11  \\
327  & 1:11:46.00 & -35:05:55.9 & 24.25$^{  0.10}_{ -0.09}$ &  24.93$^{+0.06}_{ -0.06}$ & 25.80$^{+0.14}_{ -0.13}$ &  5.19$\pm$ 0.46 &  2.41$\pm$0.21 &  49$\pm$6\\
 246  & 1:11:46.93 & -35:06:08.2 &  25.19$^{+0.14 }_{ -0.13 }$ &  26.98$^{+ 0.25 }_{ -0.21 }$ &  27.58$^{+ 0.38 }_{ -0.29 }$ &  2.18$\pm$0.26 &  1.01$\pm$0.12 &  48$\pm$10  \\
 1381  & 1:11:45.11 & -34:59:31.4 &  24.78$^{+0.12 }_{ -0.11 }$ &  25.89$^{+ 0.13 }_{ -0.11 }$ &  26.41$^{+ 0.15 }_{ -0.14 }$ &  3.18$\pm$0.33 &  1.48$\pm$0.15 &  48$\pm$8  \\
 2105  & 1:11:40.48 & -35:00:30.4 & 24.76$^{  0.13}_{ -0.12}$ & 25.70$^{+  0.11}_{ -0.10}$ & 26.33$^{+0.22}_{ -0.19}$ &  3.24$\pm$0.37 &  1.51$\pm$ 0.17 & 46$\pm$10\\ 
 2187  & 1:11:59.42 & -35:00:39.3 &  25.40$^{+ 0.08 }_{ -0.08 }$ &  27.11$^{+ 0.15 }_{ -0.14 }$ & $>$ 28.31  &  1.80$\pm$0.13 &  0.84$\pm$0.06 &  44$\pm$6  \\
 2231  & 1:11:59.30 & -35:00:38.8 &  25.20$^{+ 0.07 }_{ -0.07 }$ &  26.91$^{+ 0.12 }_{ -0.11 }$ & $>$ 28.38  &  2.16$\pm$0.14 &  1.00$\pm$0.06 &  44$\pm$4  \\
 1540  & 1:11:46.52 & -35:03:18.5 &   24.25$^{  0.13 }_{ -0.12}$ &  25.23$^{+0.08 }_{ -0.08}$ &    25.77$^{+  0.20 }_{ -0.17}$ &  5.19$\pm$ 0.59 &  2.41$\pm$ 0.27 &    43$\pm$8      \\
 2089  & 1:11:42.47 & -35:00:25.4 &  25.94$^{+0.11 }_{ -0.10 }$ &  27.36$^{+ 0.14 }_{ -0.12 }$ &  27.51$^{+ 0.22 }_{ -0.19 }$ &  1.09$\pm$0.11 &  0.51$\pm$0.05 &  40$\pm$8  \\
 1514  & 1:11:52.17 & -35:03:15.6 & 24.48$^{  0.12}_{ -0.12}$ &  24.85$^{+0.06}_{ -0.06 }$ & 25.81$^{+0.14}_{ -0.13}$ &  4.20$\pm$0.44 &  1.95$\pm$0.20 & 38$\pm$5\\ 
1982  & 1:11:37.44 & -35:00:13.1 &  25.83 $^{+ 0.15 }_{ -0.14 }$ &  27.4$^{+ 0.20 }_{ -0.17 }$ & $>$ 28.39  &  1.21$\pm$0.16 &  0.56$\pm$0.07 &  37$\pm$7  \\
 2633  & 1:11:32.45 & -35:01:34.1 &  25.90$^{+0.13 }_{ -0.12 }$ &  27.38$^{+ 0.15 }_{ -0.13 }$ &  27.7$^{+ 0.28 }_{ -0.23 }$ &  1.13$\pm$0.13 &  0.53$\pm$0.06 &  33$\pm$5  \\
 2342  & 1:11:43.93 & -35:00:56.3 &  26.24$^{+ 0.19 }_{ -0.17 }$ &  27.7$^{+ 0.25 }_{ -0.21 }$ & $>$ 28.39  &  0.83$\pm$0.13 &  0.39$\pm$0.06 &  33$\pm$8  \\
 2839  & 1:11:45.94 & -35:02:04.2 &  26.95$^{+ 0.27 }_{ -0.23 }$ &  28.38$^{+ 0.44 }_{ -0.32 }$ & $>$ 28.48  &  0.43$\pm$0.10 &  0.20$\pm$0.04 &  32$\pm$12  \\
 90  & 1:11:53.63 & -35:06:26.9 &  26.74$^{+ 0.23 }_{ -0.2 }$ &  28.16$^{+ 0.26 }_{ -0.21 }$ & $>$ 28.68  &  0.52$\pm$0.10 &  0.24$\pm$0.05 &  31$\pm$8  \\
 1680  & 1:11:55.75 & -35:03:36.2 &  26.08$^{+0.31 }_{ -0.25 }$ &  27.46$^{+ 0.32 }_{ -0.25 }$ &  27.25 $^{+ 0.25 }_{ -0.21 }$ &  0.96$\pm$0.24 &  0.45$\pm$0.11 &  29$\pm$10  \\
 100  & 1:11:28.09 & -35:06:25.9 &  26.45$^{+ 0.23 }_{ -0.20 }$ &  27.83$^{+ 0.37 }_{ -0.28 }$ & $>$ 27.37  &  0.68$\pm$0.13 &  0.32$\pm$0.06 &  29$\pm$10  \\
 1572  & 1:11:47.06 & -35:03:23.6 &  25.72$^{+0.12 }_{ -0.11 }$ &  27.02$^{+ 0.16 }_{ -0.14 }$ &  27.64$^{+ 0.37 }_{ -0.29 }$ &  1.34$\pm$0.14 &  0.62$\pm$0.07 &  27$\pm$4  \\
 2478  & 1:11:37.99 & -35:01:13.5 &  25.51$^{+0.10 }_{ -0.09 }$ &  26.81$^{+ 0.11 }_{ -0.11 }$ &  27.7$^{+ 0.36 }_{ -0.28 }$ &  1.63$\pm$0.14 &  0.76$\pm$0.07 &  26$\pm$3  \\
 2809  & 1:11:33.74 & -35:02:00.2 &  25.33$^{+0.23 }_{ -0.20 }$ &  26.56$^{+ 0.18 }_{ -0.16 }$ &  27.47$^{+ 0.34 }_{ -0.27 }$ &  1.92$\pm$0.37 &  0.89$\pm$0.17 &  24$\pm$5  \\

\end{supertabular}
\end{center}

\section{Radiative Transfer Simulations}

In this section, we briefly describe the hydrodynamical and Radiative Transfer (RT) simulations
 used throughout the paper. A detailed description will be provided in a separate paper (Cantalupo \&  Haehnelt, in preparation).

 We have performed cosmological Adaptive Mesh Refinement (AMR), hydrodynamical simulations using the publicly available 
code Ramses, version 3.03 (Teyssier 2002). The computational domain is composed by a series of four nested boxes 
at increasing, root-grid resolution. The largest box has a linear size of 40 comoving Mpc, corresponding roughly to the FWHM of the NB filter used in our 
FORS2 observation at $z=2.4$, while the high-resolution box has 
a linear size of 10 comoving Mpc, corresponding roughly to the FOV of our FORS2 observation at $z=2.4$. 
The equivalent base-grid resolution in the high-resolution region corresponds to a ($1024^3$) grid with a dark-matter 
particle mass of about $1.8\times10^6 M_{\odot}$.
 We allowed 6 additional refinements level, reaching a maximum spatial resolution of about 0.6 comoving kpc, i.e. about 180 proper pc at $z=2.4$.
The box has been centred on the most massive halo at $z=2.4$, $M_{\mathrm{DM}}\sim10^{13} M_{\odot}$, as a representative host 
of a luminous quasar. We include photo-ionization from the cosmic UV background (Haardt \& Madau 2012) using a self-shielding above 
a critical density as derived by Schaye (2001).
In each cell, gas is converted into star particles following a Schmidt law with a density threshold of 1 atom cm$^{-3}$
and a star formation timescale of 3 Gyr (see, e.g. Dubois \& Teyssier 2008). We include metal cooling, metal enrichment and supernova feedback
as described in Dubois \& Teyssier 2008.  

To produce the ``quasar-on'' models, the results of the hydrodynamical simulation have been post-processed with our RT code RADAMESH (Cantalupo \& Porciani 2011) to include 
the time-dependent ionization effects due a bright quasar with Lyman Limit luminosity of $L_{\mathrm{ll}}=5\times10^{31}$ erg s$^{-1}$ and spectral slope of $\alpha=-1.7$
 (Telfer et al. 2002). We include hydrogen and helium ionization and we evolved the temperature of the gas consistently with the ionization state using
a large number of frequency bins (50) for the ionizing radiation. 
We assumed a quasar age of 20Myr, large enough to illuminate the whole simulation box. 

In order to obtain simulated Ly$\alpha$ images we post-processed the ``quasar-off'' and ``quasar-on'' runs with our three-dimensional Ly$\alpha$ Monte Carlo RT (Cantalupo et al. 2005)
including the gas velocity field and inhomogeneous temperature distribution.
As a source of Ly$\alpha$ radiation we included HII recombination and internal star formation,
while we have decided to exclude contribution from HI collisional excitations.
 This is due to the fact that 
a proper modeling of this emission for dense systems would have required the coupling of hydro and RT, given its strong temperature dependence for a photo-ionized gas (e.g., Cantalupo et al. 2008; see also, e.g., Furlanetto et al. 2005, Faucher-Gigu\`ere et al. 2010 and Rosdahl \& Blaizot 2011 for further 
discussion), currently beyond the capabilities of the large majority of numerical studies.
 We note however that the contribution from HI collisional excitations
is likely negligible if the gas is highly ionized, as expected in proximity of an ultraluminous quasar. Partially ionized systems, e.g., optically thick clouds to the quasar radiation could present instead significant contribution from HI collisional excitations. For these sources, the simulated
Ly$\alpha$ emission, in absence of dust, needs to be considered a conservative estimate.
 
We computed the Ly$\alpha$ emissivity from the simulated Star Formation Rate (SFR)
converting the H$\alpha$-SFR relation of Kennicutt (1998) to a Ly$\alpha$-SFR relation using the Case B line ratio (e.g., Osterbrock 1989), obtaining:
SFR$=9.1\times10^{-43} L_{\mathrm{ly\alpha}}$ erg s$^{-1}$. We do not include dust absorption, likely important only for the most massive Ly$\alpha$ emitters
powered by internal star-formation, while the ``typical'' Ly$\alpha$ emitter seems to contain very little dust (e.g., Guaita et 2011). Nevertheless, continuum-selected
samples of star forming galaxies at $z\sim2.5$ (e.g., Steidel et al.2011) show that only about half of the galaxies present Ly$\alpha$ in emission 
and only 20\% of them have EW$_0>20\mathrm{\AA}$, the standard selection cut.
In other words, not every star-forming galaxy is a Ly$\alpha$ emitter and we must take this into account when deriving the simulated Luminosity Functions. 
In particular, when we computed the simulated Luminosity Function in the ``quasar-off'' model (where Ly$\alpha$ is only powered by star formation), we have used a normalization factor $n_{\mathrm{ly\alpha}}=0.5/3.7=0.13$, 
where the factor $3.7$ is necessary to renormalize 
for the overdensity of the simulated high-resolution region. Note that, given the large uncertainties related to the Ly$\alpha$ fraction and its evolution with galaxy luminosity (e.g., Stark et al. 2010), $n_{\mathrm{ly\alpha}}$ is essentially a free parameter in our model. It is remarkably, however, that with the adopted value we are able to obtain a simulated LF very
similar to the observed LF by Grove et al. (2009) and Hayes et al. (2010). In the ``quasar-on'' simulation we use the same rescaling factor, since we are interested in
the change of slope rather than normalization.
Also in this case, however, the adopted value of $n_{\mathrm{ly\alpha}}$ - combined with the steepening of the LF -
 seems to produce the right amount of sources, regardless of the possible difference in 
overdensity between quasar and random fields. Note that in the ``quasar-on'' simulation we have assumed isotropic quasar emission, while in reality
only a fraction of the volume may be illuminated by the quasar. This parameter would be degenerate with field overdensity. 
In other words, the adopted value of $n_{\mathrm{ly\alpha}}$ for the fluorescent sources 
may also be consistent with a cosmic volume 4 times more dense than a random field, if the quasar is only ``illuminating'' one fourth of such volume.

\label{lastpage}

\end{document}